%% file: arxiv.tex
\documentclass[10pt,letterpaper]{article}
\usepackage{arxiv-format}
\input{math_commands.tex}

\usepackage{bbm}
\usepackage{booktabs}
\usepackage{graphicx}
\usepackage{wrapfig}
\usepackage[table]{xcolor}
\usepackage{amsmath,amssymb,amsthm}
\usepackage[hidelinks]{hyperref}
\usepackage{url}

\theoremstyle{plain}

\title{Beyond Semantic Narrowing: Robust and Efficient LLM Watermarking with Hamming Neighborhoods}
\input{authors}
\date{}
\hypersetup{
  pdftitle={Beyond Semantic Narrowing: Robust and Efficient LLM Watermarking with Hamming Neighborhoods},
  pdfauthor={Zewen Sun, Tongyang Zhao, Liyao Xiang, Mingxuan Ma, Lingzhe Wang, Zhiyuan Li}
}

\begin{document}

\maketitle

\begin{abstract}
Semantic watermarking improves robustness against watermark removal attacks by
embedding detectable signals into sentence-level representations. However,
existing watermarking methods typically impose watermark-specific semantic preferences on
generated sentences without explicitly accounting for the highly non-uniform
and context-dependent semantic preference of LLM generation. When these
two preferences are poorly aligned, many natural
continuations become incompatible with the watermark, causing
\emph{semantic narrowing}: reduced semantic freedom, increased resampling
cost, and potential degradation on tasks with strict semantic requirements.
To alleviate this problem, we propose \emph{HammingMark}, which uses the semantic hash of the preceding sentence as a dynamic center and accepts candidates whose hashes fall within its Hamming neighborhood. Defining watermark validity over a Hamming neighborhood in compact hash space retains a larger fraction of naturally likely semantic continuations. The coarse many-to-one hash mapping further allows diverse semantic realizations to remain watermark-valid.
Experiments on C4 and BookSum show that HammingMark achieves strong robustness, high detectability, and near-unwatermarked generation quality, requiring only 2.2 sampled candidates per accepted sentence—a 72.8\% reduction compared with the most sampling-efficient existing method.
 On more complex tasks with strict semantic constraints, HammingMark achieves the highest detection rates with the highest or tied-highest ROUGE-L scores, demonstrating its effectiveness in balancing watermark detectability and generation quality under constrained generation settings.
\end{abstract}

\input{sections/1_introduction}

\input{sections/2_related_work}

\input{sections/3_method}

\input{sections/4_experiments}

\input{sections/5_conclusion}

\subsection*{AI use statement}

In this work, generative AI tools were used to assist with language polishing, manuscript drafting and revision, formatting, and literature search. All AI-assisted outputs, including retrieved information and suggested text, were critically reviewed, verified, and revised by the authors. The authors take full responsibility for the accuracy, integrity, and final content of the paper.
We take responsibility for the final content of this work, including any text or artifacts produced with the aid of generative AI.

\subsection*{Ethics statement}

This work does not involve human-subject studies or the collection of personally identifiable or sensitive information. Our experiments are conducted using publicly available or appropriately licensed data and standard evaluation protocols. We do not intend the proposed methods to facilitate harmful, discriminatory, privacy-invasive, or otherwise unethical applications. Nevertheless, as with many machine learning methods, the proposed approach may inherit biases or limitations from the data and models on which it relies, and we encourage careful evaluation before deployment in safety- or socially sensitive settings. We have made our best effort to ensure research integrity, appropriate attribution, and compliance with relevant ethical and legal requirements. The authors declare no conflicts of interest that would affect the presentation or interpretation of the results.

\subsection*{Reproducibility statement}

We have made efforts to ensure the reproducibility of our results. The appendix provides the complete experimental setup and all details necessary to reproduce our experiments, including implementation details, evaluation configurations, hyperparameter settings, data processing procedures, and other relevant experimental specifications. We also provide additional results and analyses where appropriate to facilitate verification of the reported findings. Together, these materials are intended to enable independent reproduction of the main results presented in this work. The source code will be made publicly available upon acceptance of the paper.

\bibliography{references}
\bibliographystyle{references}

\appendix

\input{sections/appendix}

\end{document}

%% file: math_commands.tex
\usepackage{amsmath,amsfonts,bm}

\def\eqref#1{equation~\ref{#1}}

\def\1{\bm{1}}

\DeclareMathAlphabet{\mathsfit}{\encodingdefault}{\sfdefault}{m}{sl}
\SetMathAlphabet{\mathsfit}{bold}{\encodingdefault}{\sfdefault}{bx}{n}



%% file: authors.tex
\author{%
  Zewen Sun\textsuperscript{1,2}\quad
  Tongyang Zhao\textsuperscript{3}\quad
  Liyao Xiang\textsuperscript{1,2}\thanks{Corresponding author.}\\[0.3em]
  Mingxuan Ma\textsuperscript{1}\quad
  Lingzhe Wang\textsuperscript{1}\quad
  Zhiyuan Li\textsuperscript{1}\\[0.5em]
  {\small \textsuperscript{1}Shanghai Jiao Tong University}\\
  {\small \textsuperscript{2}Shanghai Innovation Institute}\\
  {\small \textsuperscript{3}Northwest Polytechnical University, Xi'an}\\[0.4em]
  {\small\texttt{zwsun@sjtu.edu.cn}\quad
  \texttt{zhaotongyang@mail.nwpu.edu.cn}\quad
  \texttt{xiangliyao08@sjtu.edu.cn}}\\
  {\small\texttt{ru.jiang@sjtu.edu.cn}\quad
  \texttt{wanglingzhe@sjtu.edu.cn}\quad
  \texttt{willmaths@sjtu.edu.cn}}
}

%% file: sections/1_introduction.tex
\section{Introduction}

Large language models (LLMs) have demonstrated remarkable capabilities in generating fluent and coherent text, which simultaneously raises growing concerns about content provenance, misuse, and copyright protection \citep{liang2026watermarking, xu2025copyright, casper2026open, yang2025watermarking}. Text watermarking provides a promising solution by embedding statistically detectable signals into model-generated content without altering its intended functionality \citep{kirchenbauer2023watermark, dathathri2024scalable, liu2026distilling}. However, conventional token-level watermarking methods encode such signals by manipulating token distributions and are therefore vulnerable to paraphrasing, synonym substitution, and other semantic-preserving attacks \citep{lu2024entropy, hu2024unbiased, sun2026watermoe}. Semantic-level watermarking has recently emerged as a more robust alternative against semantic-preserving attacks by shifting the sentence, rather than the token distributions \citep{liu2024semantic, hou2024semstamp, ren2024robust}, hence is more attractive to reliable provenance verification of LLM-generated text.

Despite their improved robustness, existing semantic watermarking methods inevitably experience \emph{semantic narrowing}: they typically sample multiple candidate sentences until finding one that falls into a predefined semantic subspace. For example, bucket-based methods require the generated sentence to fall into a designated semantic region \citep{hou2024semstamp, hou2024k}, while similarity-based methods constrain its embedding to a predefined similarity interval \citep{dabiriaghdam2025simmark,zhang2025cohemark}. Multi-dimensional semantic partitioning methods similarly favor sentences located in watermark-specific regions of the embedding space \citep{huo2026pmark}. Although implemented differently, these methods impose a watermark-specific
semantic preference that may not align with the model's natural
generation preference. Poor alignment can reject otherwise plausible,
high-probability continuations, leading to inefficient, repeated sampling and potential
semantic distortion. This problem is particularly severe in constrained tasks, where the set of acceptable continuations is already limited.

Our core observation is that the natural next-sentence generation is typically not uniformly distributed over the semantic space, but puts much weights on a context-dependent set of plausible semantic
continuations \citep{meister2023locally, farquhar2024detecting, aichberger2025improving}. 
Previous semantic watermarking methods typically predefine a randomly positioned semantic subspace as the watermark-valid region, without considering whether this subspace aligns with the context-dependent semantic space favored by natural next-sentence generation.


Motivated by this principle, we propose \emph{HammingMark}, which uses the semantics of the preceding sentence to dynamically determine which next-sentence candidates are compatible with the watermark. Rather than steering generation toward a fixed, context-independent semantic region, HammingMark adapts the watermark constraint to the local context already established by the generated text, making it more likely to preserve continuations that the model naturally prefers. Within this context-aligned design, watermark selectivity operates on coarse semantic hash codes, whose many-to-one mapping weakens the coupling between watermark validity and specific semantic content. The constraint thus remains selective without prescribing a single semantic direction, allowing multiple semantically distinct continuations to remain valid and preserving flexibility for discourse changes such as elaboration, contrast, and topic shifts.


Our contributions are threefold: 1) \textbf{A unified analysis of semantic narrowing.} We provide the first systematic analysis of \emph{semantic narrowing}, a common limitation of existing semantic watermarking methods.
2) \textbf{A local-context-aligned semantic acceptance constraint for watermark embedding.} We propose \emph{HammingMark}, which centers the Hamming acceptance neighborhood on the semantic code of the preceding sentence, allowing the watermark constraint to better align with naturally plausible continuations while preserving semantic flexibility. 3) \textbf{High-fidelity, efficient, and robust semantic watermarking.} Experiments on open-ended and constrained generation tasks show that HammingMark maintains strong detectability and robustness while substantially reducing semantic restriction and resampling overhead, requiring only 2.2 sampled candidates per accepted sentence—a 72.8\% reduction compared with the most sampling-efficient existing method.


%% file: sections/2_related_work.tex
\section{Related Work}
\label{sec:related_work}


SemStamp \citep{hou2024semstamp} partitions the semantic space using locality-sensitive hashing, while K-SemStamp \citep{hou2024k} uses clustering-based regions. Both restrict generation to selected valid regions without accounting for the nonuniform, context-dependent distribution of natural continuations. These regions may therefore exclude much of the probability mass assigned to plausible continuations, causing semantic narrowing through misalignment with natural generation. SimMark \citep{dabiriaghdam2025simmark} considers inter-sentence relations but requires consecutive sentences to satisfy a predefined similarity interval. However, natural discourse does not maintain a uniform similarity range: contrast, elaboration, and topic shifts may require continuations outside this interval, which the constraint consequently rejects. This interval-based criterion can also be sensitive to paraphrasing, as changes to both sentences may shift their measured similarity across the acceptance boundary even when their meanings are largely preserved.

PMark \citep{huo2026pmark} introduces a proxy-function-based framework with multi-channel constraints. Its online variant offers strong theoretical guarantees through a distortion-free construction, but its substantial verification cost and demanding requirements for generator access, reproducible inference, and sentence-position alignment severely limit its practical applicability. Specifically, verification requires repeated sentence-level generation to reconstruct channel-wise medians, whose estimates may shift when prompts, context, or inference settings differ. Moreover, the strict sentence–key alignment requirement can render detection infeasible when sentence truncation or missing prompts disrupt this correspondence.
 A detailed discussion is provided in Appendix~\ref{app:online_pmark}.
 Its offline variant replaces dynamically estimated channel-wise medians with a fixed zero prior, eliminating generator queries but losing the exact distortion-free guarantee. Rather than rejecting candidates against an acceptance threshold, it selects from a sampled pool the candidate with the highest agreement with prescribed channel-wise random signs. The random projection matrix and target signs jointly impose fixed preferences in semantic space, concentrating selection probability on candidates favored by these preferences. Consequently, naturally plausible but lower-scoring continuations are systematically disadvantaged, inducing semantic narrowing even without explicit rejection sampling.

HammingMark defines watermark validity through a Hamming neighborhood centered on the preceding sentence’s hash code, aligning the constraint with continuations naturally favored by the local context. Meanwhile, the coarse many-to-one mapping weakens the coupling between watermark validity and specific semantic content, allowing semantically distinct continuations to satisfy the same code-level constraint. Together, these mechanisms reduce semantic narrowing while preserving flexibility for contrast, elaboration, and topic shifts. A detailed analysis of semantic narrowing across methods is provided in Appendix~\ref{app:narrowing_effectiveness} and ~\ref{app:pmark_narrowing}.

%% file: sections/3_method.tex
\section{Method}
\label{sec:method}

\begin{figure}[t]
    \centering
    \includegraphics[width=\linewidth]{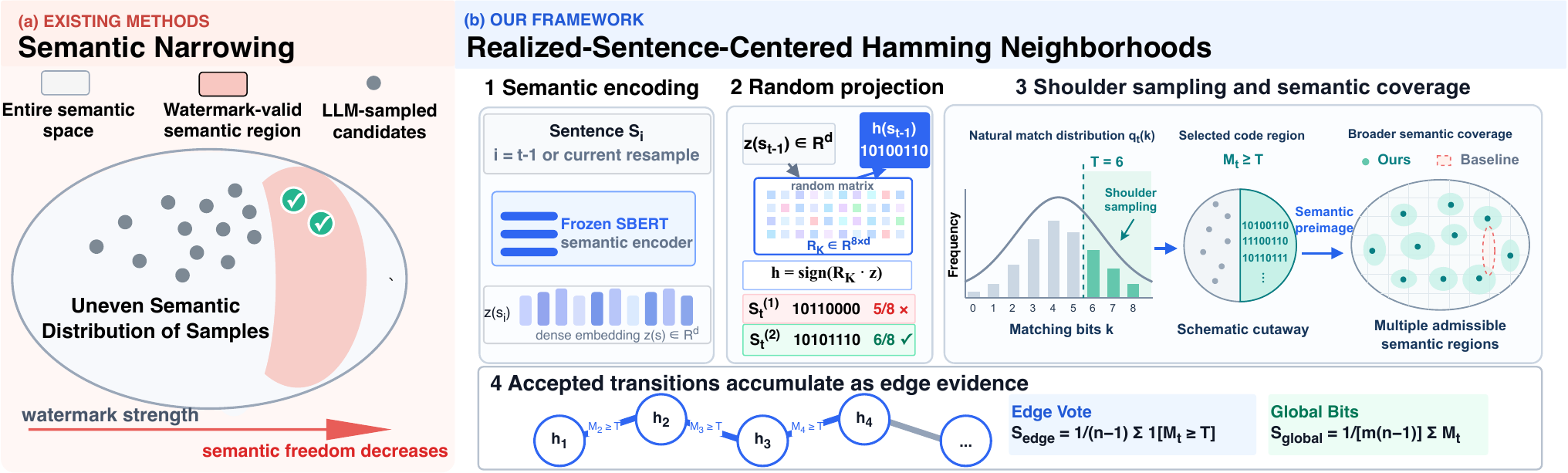}
    \caption{
    \textbf{Overview of HammingMark.}
    (a) Existing semantic watermarks may cause semantic narrowing when predefined watermark regions poorly overlap with natural generation.
(b) HammingMark uses transition-centered Hamming neighborhoods for better alignment with natural generation, while coarse many-to-one hashing preserves diverse semantic transitions.
    }
    \label{fig:framework}
\end{figure}

\subsection{Semantic Narrowing under Watermark Constraints}
\label{sec:semantic_narrowing}

We characterize semantic narrowing in terms of the natural probability mass excluded by a watermark constraint.

At generation step $t$, let $c_t$ denote the preceding context, and let $S_t \sim P_t(\cdot \mid c_t)$ denote a semantic continuation drawn from the context-dependent natural generation distribution. Here, $P_t(\cdot \mid c_t)$ captures the distribution of semantic continuations that the underlying language model would naturally produce given $c_t$. Let $W_t$ denote the set of semantic continuations that satisfy the watermark constraint at step $t$.

We define the \emph{natural acceptance mass} of the watermark constraint as
\begin{equation}
    \alpha_t
    =
    \Pr_{S_t \sim P_t(\cdot \mid c_t)}
    \left[
        S_t \in W_t
    \right].
    \label{eq:natural_acceptance}
\end{equation}

The quantity $\alpha_t$ measures how much probability mass of the natural generation distribution remains compatible with the watermark constraint. Correspondingly, the excluded natural probability mass is $1-\alpha_t$, which quantifies the degree of semantic narrowing induced by the watermark. A smaller $\alpha_t$ indicates poorer alignment between the watermark constraint and the natural distribution, causing more naturally likely continuations to be excluded.

\subsection{HammingMark: Adaptive Hamming-Neighborhood Acceptance}
\label{sec:hammingmark}

We construct a context-dependent semantic acceptance rule based on transitions between compact hash codes. For each sentence $S_t$, we obtain a semantic embedding $e_t = F(S_t) \in \mathbb{R}^{d}$ and project it using a fixed random-hyperplane matrix $R \in \mathbb{R}^{m \times d}$. The resulting semantic hash is
\begin{equation}
    h_t
    =
    H(e_t)
    =
    \mathbf{1}[Re_t \ge 0],
    \qquad
    h_t \in \{0,1\}^{m}.
    \label{eq:semantic_hash}
\end{equation}
Random-hyperplane hashing preserves angular locality in probability: semantically similar embeddings are more likely to produce matching hash bits, providing a discrete but locality-sensitive representation of sentence semantics. A detailed analysis of this locality-preserving property is provided in Appendix~\ref{app:hash_locality}.

Given the preceding code $h_{t-1}$, we define the number of matching bits between two consecutive semantic hashes as $\operatorname{Match}(h_{t-1},h_t) = m - d_H(h_{t-1},h_t)$, where $d_H(\cdot,\cdot)$ denotes Hamming distance. HammingMark accepts the current sentence when
\begin{equation}
    \operatorname{Match}(h_{t-1},h_t) \ge T
    \quad\Longleftrightarrow\quad
    d_H(h_{t-1},h_t) \le r,
    \qquad
    r = m-T.
    \label{eq:hamming_accept}
\end{equation}
Accordingly, the valid next codes form the Hamming neighborhood
\begin{equation}
    \mathcal{B}_r(h_{t-1})
    =
    \left\{
        h \in \{0,1\}^{m}
        :
        d_H(h,h_{t-1}) \le r
    \right\}.
    \label{eq:hamming_ball}
\end{equation}

Under the general formulation in Section~\ref{sec:semantic_narrowing}, the natural acceptance mass of HammingMark is
\begin{equation}
    \alpha_t^{\mathrm{HM}}
    =
    \Pr_{S_t \sim P_t(\cdot \mid c_t)}
    \left[
        H(F(S_t)) \in \mathcal{B}_r(h_{t-1})
    \right].
    \label{eq:hm_natural_acceptance}
\end{equation}
During watermark generation, a candidate sentence $S_t$ is accepted if
$\operatorname{Match}(h_{t-1},h_t) \ge T$ and is resampled otherwise. Thus,
$\alpha_t^{\mathrm{HM}}$ directly measures the natural probability mass retained by the HammingMark acceptance constraint.

The preceding hash $h_{t-1}$ has two roles. First, it determines the center of the current Hamming acceptance constraint, making the watermark rule dependent on the realized preceding semantics rather than on an independently specified target. Second, it serves as one endpoint of the inter-sentence relation statistic used by the detector, so that the same pairwise semantic relation governs both watermark insertion and verification. HammingMark still performs semantic selection by rejecting candidates that violate~\eqref{eq:hamming_accept}; its advantage lies in how the acceptance constraint is placed and represented.

\subsection{Why HammingMark Alleviates Semantic Narrowing}
\label{sec:tradeoff_alleviation}

Under the formulation in Section~\ref{sec:semantic_narrowing}, HammingMark alleviates semantic narrowing through two complementary mechanisms. First, the realized-sentence-centered Hamming constraint exploits the semantic continuity of natural sentence transitions through context-aligned shoulder sampling. Second, the coarse many-to-one hash representation allows semantically different continuations to remain valid under the same acceptance constraint.

\paragraph{Context-aligned shoulder sampling.}
As described in Section~\ref{sec:hammingmark}, HammingMark centers its acceptance constraint on the semantic hash of the preceding sentence. The key observation is that natural adjacent sentences typically exhibit semantic continuity, and random-hyperplane hashing preserves this locality in probability.

For a naturally sampled continuation $S_t$, define the natural match distribution and the corresponding HammingMark acceptance mass as
\begin{equation}
    q_t(k)
    =
    \Pr_{S_t \sim P_t(\cdot \mid c_t)}
    \left[
        \operatorname{Match}\bigl(h_{t-1},H(F(S_t))\bigr)=k
    \right],
    \qquad
    \alpha_t^{\mathrm{HM}}
    =
    \sum_{k=T}^{m} q_t(k).
    \label{eq:natural_match_acceptance}
\end{equation}

To characterize this distribution, let $e_{t-1}$ and $e_t$ be normalized sentence embeddings with cosine similarity $\rho=e_{t-1}^{\top}e_t$. Under random-hyperplane hashing,
\begin{equation}
    \operatorname{Match}(h_{t-1},h_t)\mid\rho
    \sim
    \operatorname{Binomial}
    \left(
        m,
        1-\frac{\arccos(\rho)}{\pi}
    \right).
    \label{eq:match_binomial}
\end{equation}
Therefore, $\Pr[\operatorname{Match}(h_{t-1},h_t)\ge T\mid\rho]$ increases monotonically with semantic similarity. Since natural adjacent sentences already tend to exhibit semantic continuity, their match distribution is biased toward moderate or high values. HammingMark thus performs a mild \emph{shoulder sampling} over this existing distribution rather than directing generation toward an unrelated semantic target. Importantly, this shoulder truncation operates in the discrete code space rather than imposing a hard cutoff in the underlying semantic space.

For example, with $m=8$ and $T=6$, candidates producing four, five, or six matching bits are already relatively close in hash similarity. Selecting candidates with six or more matching bits therefore introduces a local preference along an existing natural transition tendency, rather than imposing a qualitatively different semantic direction.

\paragraph{Semantic diversity under coarse hashing.}
High natural acceptance alone does not guarantee that watermark-valid continuations are semantically diverse. HammingMark additionally exploits the many-to-one nature of compact semantic hashing. The mapping $H:\mathcal{Z}\rightarrow\{0,1\}^{m}$ compresses a high-dimensional semantic representation into a short binary code, so multiple distinct semantic representations may map to the same hash code.

For two normalized semantic representations $z$ and $z'$ with angular distance $\theta$, random-hyperplane hashing gives
\begin{equation}
    d_H\!\left(H(z),H(z')\right)
    \sim
    \operatorname{Binomial}
    \left(
        m,
        \frac{\theta}{\pi}
    \right),
    \label{eq:hamming_binomial}
\end{equation}
and therefore
\begin{equation}
    \Pr\!\left[
        d_H(H(z),H(z')) \le r
    \right]
    =
    \sum_{j=0}^{r}
    {m\choose j}
    \left(\frac{\theta}{\pi}\right)^j
    \left(1-\frac{\theta}{\pi}\right)^{m-j}.
    \label{eq:hamming_probability}
\end{equation}

Semantic proximity increases the probability of nearby hash codes, but the converse need not hold for a short code. Since the compact hash provides only a sparse and coarse representation of the underlying semantic space, a higher bit match does not necessarily imply greater semantic similarity than a lower bit match. Importantly,~\eqref{eq:hamming_probability} shows that even when two semantic representations are less similar, there can still be a nonzero probability that their hash codes fall within the same Hamming neighborhood. Therefore, although HammingMark performs shoulder sampling in the hash space, it does not explicitly filter out continuations solely because they have lower semantic similarity. Since the mapping is many-to-one, semantically different continuations may share the same code or map to different codes that are simultaneously accepted by the Hamming constraint. Watermark validity therefore does not correspond to a single semantic realization or direction. This flexibility is particularly useful for contrast, elaboration, topic shifts, and the introduction of new information, where a continuation may differ substantially from the preceding sentence while still satisfying the watermark rule. Appendix~\ref{app:semantic_diversity} provides a formal analysis of this many-to-one property.

Together, these two mechanisms address complementary sources of semantic restriction. Context-aligned shoulder sampling makes watermark selection follow an existing tendency of natural semantic transitions rather than imposing an unrelated semantic target, while coarse many-to-one hashing preserves semantic diversity among watermark-valid continuations.

\subsection{Sequence-Level Detection}
\label{sec:detection}

Given a text containing $n$ sentences $S_1,\ldots,S_n$, the detector recovers
their semantic hashes and adjacent matching scores as
\begin{equation}
    h_t=H(F(S_t)),
    \qquad
    M_t=\operatorname{Match}(h_{t-1},h_t),
    \quad t=2,\ldots,n.
    \label{eq:detection_scores}
\end{equation}
We use two complementary sequence-level statistics.

\paragraph{Edge Vote.}
Each adjacent sentence pair is treated as one transition unit. A transition
casts a valid vote when $M_t\ge T$, and the final score is
\begin{equation}
    S_{\mathrm{edge}}
    =
    \frac{1}{n-1}
    \sum_{t=2}^{n}
    \mathbf{1}[M_t\ge T].
    \label{eq:edge_vote}
\end{equation}

\paragraph{Global Bits.}
Instead of binarizing each transition, Global Bits aggregates all matching
bits,
\begin{equation}
    S_{\mathrm{global}}
    =
    \frac{1}{m(n-1)}
    \sum_{t=2}^{n}M_t.
    \label{eq:global_bits}
\end{equation}
It therefore retains partial evidence from transitions that no longer exceed
$T$ after semantic-preserving perturbations.

For either detector $d\in\{\mathrm{edge},\mathrm{global}\}$, a text is
classified as watermarked when $S_d>\tau_d$, where $\tau_d$ is calibrated
from unwatermarked text at the target false-positive rate. Detection requires
only the fixed semantic encoder and projection matrix, without access to the
generation model.

%% file: sections/4_experiments.tex
\section{Experiments}
\label{sec:experiments}

We organize our evaluation around the following research questions.
\textbf{RQ1 (Detectability):} Can HammingMark be reliably distinguished from
unwatermarked text across datasets and language models?
\textbf{RQ2 (Robustness):} Does the watermark remain detectable after
local edits, sentence-level paraphrasing, and document-level rewriting?
\textbf{RQ3 (Text quality):}
How well does HammingMark preserve the quality of generated text?
\textbf{RQ4 (Sampling efficiency and semantic overlap):}
How well does the watermark-valid region overlap with naturally preferred
continuations, and how does this affect rejection-sampling efficiency?
\textbf{RQ5 (Complex tasks):} Does HammingMark remain effective on conditional,
long-form generation tasks?
\textbf{RQ6 (Semantic sampling visualization):}
How do different semantic watermarking methods constrain naturally sampled
candidate continuations in the semantic space?


\providecommand{\expTBD}{\textit{TBD}}





\subsection{RQ1--RQ3: Detectability, Robustness, and Quality}
\label{sec:main_results}

\begin{table*}[t]
    \centering
    \caption{
        \textbf{Clean detectability, robustness, and generation quality on C4 and BookSum.}
        Each detection entry reports
        $\mathrm{TPR}@1\%/\mathrm{TPR}@5\%/\mathrm{AUROC}$.
    }
    \label{tab:main_results}
    \scriptsize
    \setlength{\tabcolsep}{2.2pt}
    \renewcommand{\arraystretch}{1.03}

    \resizebox{\textwidth}{!}{%
        \begin{tabular}{@{}lcccccc@{}}
            \toprule
            Method
             & Clean
             & Parrot
             & Pegasus
             & DIPPER
             & API
             & PPL $\downarrow$                                                     \\
            \midrule

            \multicolumn{7}{l}{\textit{Mistral-7B \citep{jiang2023mistral7b} on C4 \citep{raffel2020exploring}}}                           \\
            \rowcolor{gray!15}
            No watermark
             & --                                        & -- & -- & -- & -- & 4.51 \\

            KGW \citep{kirchenbauer2023watermark}
             & \textbf{1.00/1.00/1.00}
             & 0.59/0.72/0.90
             & 0.51/0.63/0.89
             & 0.43/0.49/0.84
             & 0.31/0.39/0.80
             & 8.00                                                                 \\

            SynthID \citep{dathathri2024scalable}
             & \textbf{1.00/1.00/1.00}
             & 0.32/0.45/0.82
             & 0.23/0.42/0.80
             & 0.35/0.40/0.80
             & 0.24/0.38/0.79
             & \textbf{4.52}                                                        \\

            MorphMark \citep{wang2025morphmark}
             & 0.99/1.00/1.00
             & 0.54/0.60/0.88
             & 0.47/0.55/0.84
             & 0.51/0.55/0.86
             & 0.53/0.64/0.89
             & 6.62                                                                 \\

            SIR \citep{liu2024semantic}
             & 0.98/1.00/1.00
             & 0.51/0.63/0.91
             & 0.56/0.60/0.90
             & 0.44/0.57/0.84
             & 0.38/0.46/0.82
             & 8.35                                                                 \\

            K-SemStamp \citep{hou2024k}
             & 0.83/0.88/0.98
             & 0.43/0.58/0.86
             & 0.46/0.65/0.89
             & 0.66/0.87/0.95
             & 0.31/0.40/0.80
             & 10.78                                                                \\

            SemStamp \citep{hou2024semstamp}
             & 0.82/0.87/0.97
             & 0.56/0.77/0.91
             & 0.79/0.90/0.97
             & 0.76/0.94/0.98
             & 0.32/0.48/0.84
             & 9.36                                                                 \\

            SimMark \citep{dabiriaghdam2025simmark}
             & 0.71/0.82/0.95
             & 0.28/0.43/0.82
             & 0.35/0.51/0.84
             & 0.23/0.41/0.80
             & 0.25/0.36/0.78
             & 8.34                                                                 \\

            PMark \citep{huo2026pmark}
             & 0.99/0.99/1.00
             & 0.94/0.97/0.99
             & 0.89/0.91/0.98
             & 0.75/0.85/0.97
             & 0.91/0.93/0.98
             & 5.13                                                                 \\

            \rowcolor{gray!35}
            HammingMark (Edge Vote)
             & 0.99/1.00/1.00
             & 0.92/0.96/0.99
             & 0.85/0.98/0.99
             & 0.89/0.97/0.99
             & 0.90/0.93/0.98
             & 4.61                                                                 \\

            \rowcolor{gray!35}
            HammingMark (Global Bits)
             & 0.98/1.00/1.00
             & \textbf{0.94}/\textbf{0.99}/\textbf{1.00}
             & \textbf{0.91}/\textbf{1.00}/\textbf{0.99}
             & \textbf{0.93}/\textbf{0.98}/\textbf{0.99}
             & \textbf{0.94}/\textbf{0.97}/\textbf{0.99}
             & 4.61                                                                 \\

            \midrule
            \multicolumn{7}{l}{\textit{Mistral-7B on BookSum \citep{kryscinski2022booksum}}}                      \\
            \rowcolor{gray!15}
            No watermark
             & --                                        & -- & -- & -- & -- & 6.20 \\

            KGW
             & \textbf{1.00}/\textbf{1.00}/\textbf{1.00}
             & 0.34/0.43/0.86
             & 0.40/0.45/0.82
             & 0.35/0.41/0.81
             & 0.16/0.23/0.64
             & 10.14                                                                   \\

            SynthID
             & \textbf{1.00}/\textbf{1.00}/\textbf{1.00}
             & 0.15/0.21/0.69
             & 0.10/0.16/0.60
             & 0.11/0.19/0.62
             & 0.13/0.18/0.60
             & 6.57                                                                   \\

            MorphMark
             & \textbf{1.00}/\textbf{1.00}/\textbf{1.00}
             & 0.45/0.52/0.83
             & 0.31/0.37/0.79
             & 0.42/0.55/0.81
             & 0.22/0.27/0.65
             & 9.33                                                                  \\

            SIR
             & \textbf{1.00}/\textbf{1.00}/\textbf{1.00}
             & 0.54/0.61/0.85
             & 0.47/0.63/0.87
             & 0.52/0.61/0.83
             & 0.14/0.25/0.65
             & 12.49                                                                   \\

            K-SemStamp
             & 0.79/0.91/0.98
             & 0.51/0.64/0.88
             & 0.56/0.69/0.90
             & 0.70/0.86/0.94
             & 0.16/0.27/0.69
             & 15.72                                                                   \\

            SemStamp
             & 0.88/0.93/0.98
             & 0.67/0.71/0.90
             & 0.71/0.78/0.90
             & 0.82/0.91/0.97
             & 0.20/0.35/0.71
             & 14.98                                                                   \\

            SimMark
             & 0.77/0.86/0.96
             & 0.29/0.52/0.81
             & 0.44/0.58/0.84
             & 0.21/0.43/0.80
             & 0.14/0.21/0.62
             & 14.25                                                                   \\

            PMark
             & 0.97/0.99/0.99
             & \textbf{0.96}/\textbf{0.98}/\textbf{0.99}
             & 0.79/0.91/0.98
             & 0.78/0.93/0.98
             & 0.71/0.84/0.94
             & 8.26                                                                   \\

            \rowcolor{gray!35}
            HammingMark (Edge Vote)
             & \textbf{1.00}/\textbf{1.00}/\textbf{1.00}
             & 0.91/0.94/0.99
             & 0.86/0.95/0.99
             & 0.91/0.94/0.98
             & 0.88/0.92/0.98
             & \textbf{6.40}                                                                 \\

            \rowcolor{gray!35}
            HammingMark (Global Bits)
             & \textbf{1.00}/\textbf{1.00}/\textbf{1.00}
             & 0.93/0.95/\textbf{0.99}
             & \textbf{0.90}/\textbf{0.99}/\textbf{0.99}
             & \textbf{0.91}/\textbf{0.94}/\textbf{0.99}
             & \textbf{0.92}/\textbf{0.95}/\textbf{0.99}
             & \textbf{6.40}                                                                 \\

            \midrule
            \multicolumn{7}{l}{\textit{Qwen2.5-7B \citep{DBLP:journals/corr/abs-2412-15115} on C4}}                           \\
            \rowcolor{gray!15}
            No watermark
             & --                                        & -- & -- & -- & -- & 5.52 \\

            KGW
             & \textbf{0.99}/\textbf{1.00}/\textbf{1.00}
             & 0.08/0.13/0.66
             & 0.08/0.11/0.67
             & 0.06/0.12/0.61
             & 0.02/0.08/0.49
             & 7.29                                                                 \\

            SynthID
             & \textbf{0.99}/\textbf{1.00}/\textbf{1.00}
             & 0.06/0.10/0.50
             & 0.02/0.04/0.47
             & 0.03/0.06/0.48
             & 0.02/0.06/0.48
             & \textbf{5.30}                                                        \\

            MorphMark
             & 0.98/1.00/1.00
             & 0.04/0.20/0.65
             & 0.06/0.17/0.64
             & 0.02/0.21/0.64
             & 0.03/0.11/0.53
             & 5.63                                                                 \\

            SIR
             & 0.97/1.00/1.00
             & 0.55/0.59/0.89
             & 0.62/0.68/0.92
             & 0.41/0.58/0.85
             & 0.34/0.40/0.79
             & 7.08                                                                 \\

            K-SemStamp
             & 0.78/0.93/0.95
             & 0.42/0.65/0.86
             & 0.49/0.73/0.89
             & 0.66/0.90/0.95
             & 0.17/0.29/0.77
             & 11.67                                                                \\

            SemStamp
             & 0.85/0.88/0.97
             & 0.59/0.82/0.91
             & 0.61/0.83/0.95
             & 0.74/0.96/0.97
             & 0.24/0.37/0.82
             & 9.46                                                                 \\

            SimMark
             & 0.72/0.81/0.94
             & 0.27/0.50/0.82
             & 0.38/0.52/0.86
             & 0.29/0.41/0.78
             & 0.31/0.46/0.83
             & 8.11                                                                 \\

            PMark
             & 0.99/1.00/0.99
             & 0.96/0.98/0.99
             & 0.85/0.91/0.96
             & 0.91/0.94/0.97
             & 0.83/0.89/0.98
             & 6.32                                                                 \\

            \rowcolor{gray!35}
            HammingMark (Edge Vote)
             & 0.98/0.99/0.99
             & 0.93/0.94/0.98
             & \textbf{0.89}/\textbf{0.93}/\textbf{0.98}
             & 0.92/0.94/0.98
             & 0.94/0.98/0.99
             & 5.38                                                                 \\

            \rowcolor{gray!35}
            HammingMark (Global Bits)
             & 0.98/0.99/0.99
             & \textbf{0.97}/\textbf{0.98}/\textbf{0.99}
             & \textbf{0.89}/\textbf{0.93}/\textbf{0.98}
             & \textbf{0.94}/\textbf{0.95}/\textbf{0.99}
             & \textbf{0.95}/\textbf{0.98}/\textbf{0.99}
             & 5.38                                                                 \\

            \midrule
            \multicolumn{7}{l}{\textit{Qwen2.5-7B on BookSum}}                      \\
            \rowcolor{gray!15}
            No watermark
             & --                                        & -- & -- & -- & -- & 4.45 \\

            KGW
             & \textbf{0.99}/\textbf{1.00}/\textbf{1.00}
             & 0.05/0.13/0.65
             & 0.05/0.19/0.62
             & 0.02/0.12/0.62
             & 0.03/0.12/0.53
             & 8.46                                                                   \\

            SynthID
             & \textbf{0.99}/\textbf{1.00}/\textbf{1.00}
             & 0.03/0.08/0.54
             & 0.05/0.12/0.56
             & 0.00/0.09/0.55
             & 0.01/0.08/0.51
             & \textbf{4.63}                                                                   \\

            MorphMark
             & 0.97/0.99/1.00
             & 0.02/0.05/0.57
             & 0.02/0.07/0.55
             & 0.02/0.06/0.56
             & 0.01/0.09/0.51
             & 7.31                                                                   \\

            SIR
             & 0.95/0.98/0.99
             & 0.54/0.63/0.84
             & 0.60/0.67/0.91
             & 0.46/0.66/0.85
             & 0.10/0.23/0.72
             & 9.76                                                                   \\

            K-SemStamp
             & 0.82/0.91/0.96
             & 0.44/0.70/0.88
             & 0.53/0.61/0.84
             & 0.68/0.75/0.90
             & 0.14/0.28/0.75
             & 10.30                                                                   \\

            SemStamp
             & 0.87/0.90/0.97
             & 0.57/0.84/0.92
             & 0.45/0.71/0.89
             & 0.70/0.78/0.91
             & 0.17/0.33/0.76
             & 11.49                                                                   \\

            SimMark
             & 0.71/0.85/0.93
             & 0.31/0.43/0.80
             & 0.36/0.54/0.83
             & 0.24/0.40/0.81
             & 0.13/0.21/0.72
             & 10.28                                                                   \\

            PMark
             & 0.95/0.99/0.99
             & \textbf{0.92}/\textbf{0.98}/\textbf{0.99}
             & 0.74/0.92/0.96
             & 0.78/0.89/0.97
             & 0.78/0.85/0.96
             & 6.39                                                                   \\

            \rowcolor{gray!35}
            HammingMark (Edge Vote)
             & \textbf{0.99}/\textbf{1.00}/\textbf{1.00}
             & 0.87/0.93/0.98
             & 0.83/0.89/0.97
             & 0.82/0.92/0.97
             & 0.89/0.94/0.98
             & 4.75                                                                 \\

            \rowcolor{gray!35}
            HammingMark (Global Bits)
             & 0.97/0.99/1.00
             & 0.93/0.95/0.99
             & \textbf{0.84}/\textbf{0.91}/\textbf{0.97}
             & \textbf{0.82}/\textbf{0.94}/\textbf{0.98}
             & \textbf{0.91}/\textbf{0.94}/\textbf{0.98}
             & 4.75                                                                 \\

            \bottomrule
        \end{tabular}
    }
\end{table*}

HammingMark maintains near-perfect clean-text detection across all model--dataset settings, with $\mathrm{TPR}@1\%$ of $0.97$--$1.00$ and AUROC of $0.99$--$1.00$. Edge Vote is particularly effective on clean text because it directly measures whether each transition satisfies the embedding rule.

Under paraphrasing and rewriting, Global Bits is generally more robust, achieving $\mathrm{TPR}@1\%$ of $0.82$--$0.97$ across all settings. Unlike Edge Vote, which discards a transition once it falls below the matching threshold, Global Bits retains partial evidence from modified transitions by aggregating all matching bits. Therefore, Edge Vote is preferable for ordinary clean-text verification, whereas Global Bits is better suited to attacked text.

HammingMark also preserves generation quality, with PPL differing from unwatermarked generation by at most $0.30$ across all settings. This advantage follows from the relational Hamming-ball constraint: instead of forcing each sentence into a fixed semantic region, HammingMark retains a large set of admissible semantic realizations. The model can therefore continue exploring plausible continuations and select content appropriate to the context while embedding a detectable transition-level signal.

\subsection{RQ4: Sampling Efficiency and Natural Acceptance}
\label{sec:sampling_efficiency}

\begin{table}[t]
\caption{
Sampling efficiency and estimated natural acceptance on C4.
Lower sampling cost and higher natural acceptance are better.
Random Code uses the same code-space coverage as HammingMark
($37/256$) but selects valid codes uniformly at random.
}
\label{tab:sampling_efficiency}
\centering
\resizebox{\linewidth}{!}{
\begin{tabular}{lcccccc}
\hline
Method
& SimMark
& SemStamp
& K-SemStamp
& PMark
& Random Code
& HammingMark \\
\hline
Samples / sent. $\downarrow$
& 8.1 & 99.3 & 13.3 & 64.0 & 14.7 & \textbf{2.2} \\
Tokens / sent. $\downarrow$
& 186.7 & 1694.4 & 246.9 & 1185.8 & 273.1 & \textbf{41.58} \\
Natural acceptance $\widehat{\bar{\alpha}}$ $\uparrow$
& 0.34 & 0.04 & 0.13 & / & 0.15 & \textbf{0.58} \\
\hline
\end{tabular}
}
\end{table}

As shown in Table~\ref{tab:sampling_efficiency}, HammingMark requires only
2.2 sampled candidates and 41.58 generated tokens for each accepted sentence.
Compared with SimMark, the most efficient baseline, HammingMark reduces the
number of sampled candidates by approximately 72.8\% and token consumption
by 77.7\%.

To examine the source of this sampling efficiency, we estimate the natural
acceptance mass defined in Section~\ref{sec:semantic_narrowing}. For each of
100 prompts, we independently sample 300 candidate continuations from the
unwatermarked model before applying any watermark constraint. These candidates
are therefore samples from the context-dependent natural generation
distribution $P_t(\cdot \mid c_t)$. We then apply each watermarking rule to
the sampled candidates and measure the fraction that would be accepted
without resampling.

For a given context $c_t$, we estimate the natural acceptance mass by
\begin{equation}
    \hat{\alpha}_t
    =
    \frac{1}{N}
    \sum_{i=1}^{N}
    \mathbf{1}
    \left[
        S_t^{(i)} \in W_t
    \right],
    \qquad
    S_t^{(i)} \sim P_t(\cdot \mid c_t).
    \label{eq:empirical_natural_acceptance}
\end{equation}
where $N=300$ is the number of independently sampled natural candidates.
We report $\widehat{\bar{\alpha}}$, obtained by averaging $\hat{\alpha}_t$
across the 100 evaluation prompts. Thus, the values reported in
Table~\ref{tab:sampling_efficiency} are direct Monte Carlo estimates of the
natural probability mass retained by each watermark constraint.
A larger natural acceptance mass means that a greater fraction of
continuations produced by the unwatermarked model already satisfies the
watermark rule. This quantity is directly related to rejection-sampling cost:
when natural candidates are accepted more frequently, fewer candidates need
to be sampled before obtaining a watermark-valid continuation.
HammingMark achieves an estimated natural acceptance of 0.58, meaning that
approximately 58\% of naturally sampled continuations satisfy its watermark
constraint without additional semantic redirection. This is consistent with
its substantially lower sampling cost.
We further evaluate the loss of semantic diversity induced by each method
in Appendix ~\ref{app:semantic_coverage} and ~\ref{app:semantic_diversity_exp}, complementing the natural
acceptance analysis with a direct assessment of diversity preservation.

\subsection{RQ5: Performance on Complex Generation Tasks}
\label{sec:complex_tasks}

\begin{table}[t]
    \caption{Performance on complex generation tasks on Mistral-7B. Detection results are
        reported as TPR@1\%, TPR@5\%, and AUROC.}
    \label{tab:complex_tasks}
    \centering
    \resizebox{\linewidth}{!}{
        \begin{tabular}{lcccccccc}
            \hline
            Method
             & \multicolumn{4}{c}{ELI5 \citep{fan2019eli5}}
             & \multicolumn{4}{c}{Multi-News \citep{fabbri2019multi}}                                                       \\
            \cline{2-9}
             & @1 $\uparrow$                  & @5 $\uparrow$ & AUC $\uparrow$ & Rouge-L $\uparrow$
             & @1 $\uparrow$                  & @5 $\uparrow$ & AUC $\uparrow$ & Rouge-L $\uparrow$ \\
            \hline
            \rowcolor{gray!15}
            No watermark
             & --                             & --            & --             & 0.34
             & --                             & --            & --             & 0.36                 \\
            \hline
            KGW
             & 0.78                           & 0.89          & 0.97           & 0.29
             & 0.21                           & 0.49          & 0.86           & 0.31               \\
            SynthID
             & 0.43                           & 0.67          & 0.92           & 0.31
             & 0.08                           & 0.17          & 0.64           & \textbf{0.36}      \\
            MorphMark
             & 0.49                           & 0.64          & 0.90           & 0.29
             & 0.26                           & 0.41          & 0.88           & 0.35               \\
            SIR
             & 0.63                           & 0.75          & 0.93           & 0.28
             & 0.21                           & 0.36          & 0.85           & 0.34               \\
            SemStamp
             & 0.22                           & 0.36          & 0.82           & 0.31
             & 0.03                           & 0.19          & 0.64           & 0.35               \\
            K-SemStamp
             & 0.13                           & 0.24          & 0.71           & 0.32
             & 0.03                           & 0.16          & 0.64           & 0.33               \\
            SimMark
             & 0.26                           & 0.41          & 0.84           & 0.31
             & 0.06                           & 0.11          & 0.62           & 0.34               \\
            PMark
             & 0.85                           & 0.91          & 0.98           & 0.29
             & 0.86                           & 0.92          & 0.99           & 0.32               \\
            \rowcolor{gray!35}
            HammingMark
             & \textbf{0.95}                  & \textbf{0.95} & \textbf{0.99}  & \textbf{0.33}
             & \textbf{0.91}                  & \textbf{1.00} & \textbf{0.99}  & \textbf{0.36}      \\
            \hline
        \end{tabular}
    }
\end{table}

As shown in Table~\ref{tab:complex_tasks}, HammingMark achieves TPR@1\% of 0.95 on ELI5 and 0.91 on MultiNews while preserving the highest or tied-highest ROUGE-L. These tasks strongly constrain generation semantics, concentrating the natural distribution $P_t(\cdot\mid c_t)$ on task-consistent continuations and leaving little room for semantic redirection. Poorly aligned watermark constraints can therefore reject many high-probability candidates, yielding a small natural acceptance mass $\alpha_t$ and frequent resampling.
HammingMark mitigates this issue through mild shoulder sampling around the realized semantic transition. Semantically coherent continuations are more likely to have high hash similarity, while the many-to-one hash mapping allows multiple semantic realizations to satisfy the same Hamming constraint. This preserves more natural probability mass, helping maintain task fidelity while accumulating a detectable watermark signal.

\subsection{RQ6: Visualizing Natural Acceptance under Semantic Constraints}
\label{sec:sampling_visualization}

\begin{figure}[t]
\centering
\includegraphics[width=\linewidth]{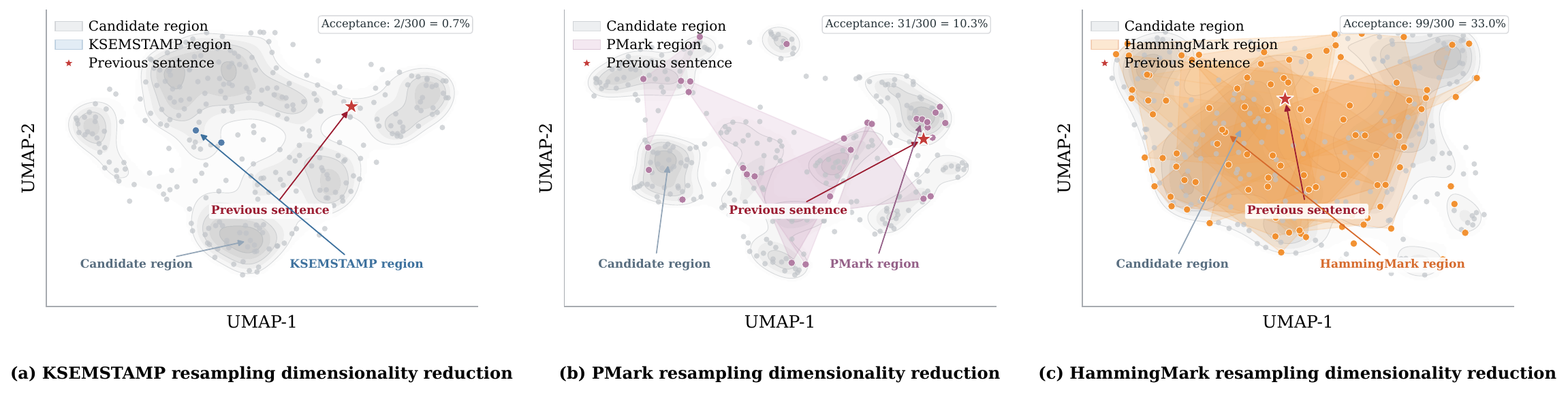}
\caption{
\textbf{Visualization of natural candidates accepted by different watermark constraints.}
UMAP projection of 300 independently sampled continuations from the
unwatermarked model for a representative challenging prompt. All methods are
compared under matched watermark effectiveness. For offline PMark, we highlight candidates matching all key-specified channel signs under the zero-median prior.
Highlighted points denote candidates accepted by the watermark
constraint.
}
\label{fig:sam_analysis}
\end{figure}

To provide an intuitive candidate-level view of natural acceptance, we independently sample 300 continuations from the unwatermarked model for a representative challenging prompt and project their sentence embeddings into two dimensions with UMAP. For each method, candidates satisfying its watermark constraint are highlighted within the same natural sample set. The highlighted fraction thus gives a prompt-level Monte Carlo estimate of the natural acceptance mass.

As shown in Figure~\ref{fig:sam_analysis}, K-SemStamp accepts only 2 of the
300 natural candidates, while PMark accepts 31 and HammingMark accepts 99.
Thus, for this prompt, HammingMark retains a substantially larger fraction of
the continuations that the unwatermarked model naturally produces. The
highlighted HammingMark candidates are also distributed across multiple parts
of the projected sample cloud, illustrating that watermark-valid candidates
are not confined to a single local semantic realization.
This visualization complements the aggregate natural-acceptance results in
Table~\ref{tab:sampling_efficiency}. Importantly, the UMAP geometry is used
only for qualitative illustration: semantic narrowing is quantified by the
fraction of natural probability mass retained by the watermark constraint,
rather than by the geometric area occupied by highlighted points in the
two-dimensional projection.

\subsection{Ablation Study}
\label{sec:ablation}

\paragraph{Effect of constraint placement.}
We examine whether HammingMark's higher natural acceptance is due simply to
the number of valid hash codes or to how these codes are placed.

With $m=8$ and $T=6$, HammingMark accepts
$\sum_{j=0}^{2}\binom{8}{j}=37$ of the $256$ possible hash codes. As a
control, we construct a Random Code baseline that also accepts 37 codes, but
selects them uniformly at random and independently of the preceding sentence.
All other components and evaluation settings remain unchanged.

As shown in Table~\ref{tab:sampling_efficiency}, Random Code achieves a natural
acceptance of $0.15$, whereas HammingMark reaches $0.58$, despite identical
code-space coverage. For a random set $\mathcal{A}$ of 37 codes, each code is
included with probability $37/256$. Therefore, for any natural hash
distribution $p_t(h)$,
\begin{equation}
    \mathbb{E}_{\mathcal{A}}
    \left[
        \alpha_t(\mathcal{A})
    \right]
    =
    \sum_h p_t(h)\Pr_{\mathcal{A}}[h\in\mathcal{A}]
    =
    \frac{37}{256}
    \approx 0.1445.
    \label{eq:random_code_expectation}
\end{equation}
The observed value of $0.15$ is close to this random-set expectation.

In contrast, HammingMark places the same number of valid codes around the
preceding sentence hash. Combined with the locality-preserving property of
random-hyperplane hashing, this transition-centered placement makes the
constraint better aligned with natural semantic transitions. The large gap
between $0.15$ and $0.58$ therefore shows that natural acceptance depends
strongly on constraint placement rather than code-space coverage alone.

%% file: sections/5_conclusion.tex
\section{Limitations}
\label{sec:limitations}
HammingMark depends on the quality of the semantic encoder and sentence
segmentation; unstable embeddings or ambiguous sentence boundaries may
alter hash codes and reduce detection reliability. Although HammingMark
substantially improves sampling efficiency, it remains a rejection-sampling method and may incur additional cost under highly restrictive contexts or strict matching thresholds.

\section{Conclusion}
\label{sec:conclusion}
This work identifies \emph{semantic narrowing} as the misalignment between watermark constraints and naturally preferred generation. We propose \emph{HammingMark}, a realized-sentence-centered semantic watermarking method that reduces this mismatch through context-aligned Hamming acceptance while preserving semantic flexibility. Across open-ended and constrained generation tasks, HammingMark achieves strong detectability and robustness with near-unwatermarked quality and low sampling cost.

%% file: sections/appendix.tex
\section{Practical Limitations of Online PMark}
\label{app:online_pmark}

PMark \citep{huo2026pmark} makes an important theoretical contribution
to semantic watermarking by introducing a proxy-function framework
with a distortion-free online construction. At each generation step,
it samples candidate sentences and recursively partitions them into
equally sized subsets using channel-wise empirical medians. Under
the prescribed random channel decisions and balanced partitioning,
each sampled candidate has the same marginal selection probability.
Averaging over these decisions and candidate sampling therefore
recovers the original next-sentence distribution. This provides a
principled foundation for semantic watermarking. However, this
generation-side guarantee does not by itself ensure inexpensive,
reproducible, or context-independent verification.

\paragraph{Generator-dependent verification and computational cost.}
Online PMark requires the detector to resample $N$ candidate
continuations for each sentence under its preceding context and
estimate channel-wise medians from these samples. For $D$ documents
containing $L$ sentences each, this entails approximately $DLN$
sentence generations, in addition to semantic encoding and evidence
aggregation. Verification therefore requires access to the original
generator or a sufficiently faithful generation service, making
large-scale screening costly and limiting deployment by independent
verifiers. Hosting the generator also introduces substantial memory
and inference requirements; relying on a remote service instead
introduces availability, latency, and version-control dependencies.

\paragraph{Sensitivity to changes in the generation distribution.}
The published detector uses fresh samples rather than requiring
exact replay of the original candidate pool. Consequently, a change
in the sampling seed alone does not invalidate the distortion-free
generation guarantee. Nevertheless, reliable median reconstruction
depends on reproducing the relevant conditional generation
distribution. Model updates or fine-tuning, changes in prompt
formatting or decoding settings, and numerical differences that
affect sampling can alter this distribution and its proxy medians.
Even under unchanged settings, finite-sample estimation introduces
discrepancies between generation and detection. PMark uses soft
counting to mitigate such discrepancies, but this mechanism does
not guarantee invariance to systematic changes in the generator or
its inference configuration.

\paragraph{Key reuse and repeated-query limitations.}
PMark explicitly acknowledges in its Appendix~D that its predefined,
position-indexed random seeds leave $n$-shot undetectability unresolved,
and restricts its guarantee to single-shot distortion-freeness
\citep{huo2026pmark}. Specifically, averaging over random watermark
decisions recovers the natural next-sentence distribution, but this
does not establish that repeated responses under a reused key follow
the joint distribution of independent unwatermarked samples.
When the same prompt is queried repeatedly with the same key sequence,
the corresponding channel preferences are repeatedly imposed rather
than independently randomized across responses. Consequently, the
single-shot guarantee does not ensure preservation of natural output
variability in repeated-query settings. This is an acknowledged
limitation of the construction: PMark leaves dynamic semantic-level
seed generation for addressing this issue to future work.

\paragraph{Dependence on prompts and sentence--key alignment.}
For sentence $s_t$, the online detector reconstructs medians from
the conditional distribution
\[
P_M(\cdot \mid \pi,s_1,\ldots,s_{t-1}),
\]
while evaluating watermark evidence with the corresponding
position-indexed channel keys and prescribed proxy directions.
This requires both the appropriate preceding context and the
correct sentence index. When a circulated text contains only the
generated response, the original prompt is unavailable, preventing
faithful reconstruction of the conditioning context. Even when
prompt text is included, an unmarked prompt--response boundary
leaves the starting key index ambiguous. Prefix truncation or
sentence insertion and deletion can further change the available
context and shift subsequent sentence--key correspondence.
These operations can occur through ordinary quotation or editing,
without substantial semantic rewriting. The published detection
procedure provides no explicit mechanism for recovering missing
context or resynchronizing unknown sentence offsets. Searching
over possible offsets would add computational cost and require
appropriate false-positive calibration, while still not restoring
missing conditioning information.

Overall, online PMark offers a valuable theoretical construction,
but its practical verification remains tightly coupled to the
generator, conditioning context, and sentence indexing used during
generation. These dependencies substantially limit its suitability
for scalable detection of independently circulated text.

\section{Semantic Narrowing and Watermark Effectiveness}
\label{app:narrowing_effectiveness}

\paragraph{Natural acceptance mass.}
Fix a generation context $c_t$ and let $z$ denote a normalized sentence
embedding with natural distribution $p_t(z)=p(z\mid c_t)$.
For a rejection-sampling watermark, let $a_t(z)\in\{0,1\}$ be its
acceptance rule under a fixed key and context. Define
\begin{equation}
\alpha_t=\int p_t(z)a_t(z)\,dz,
\qquad N_t=1-\alpha_t.
\label{eq:app_acceptance_mass}
\end{equation}
Integrals are understood with respect to the embedding distribution,
including discrete distributions. The excluded natural probability mass
$N_t$ is an operational measure of semantic narrowing; it is
not the proportional decrease in semantic diversity, entropy, or generation
quality. For $\alpha_t>0$, ideal rejection sampling with unlimited budget
induces
\begin{equation}
p_t^{\mathrm{WM}}(z)=\frac{p_t(z)a_t(z)}{\alpha_t}.
\end{equation}
We initially omit robustness margins and fallback, and introduce a finite
budget below. Each method may use a different encoder and thus a different
embedding distribution. Geometric reference calculations therefore cannot
replace comparisons on common candidate texts from shared contexts.

\subsection{Semantic Narrowing in Partition- and Similarity-Based Methods}

\paragraph{SemStamp and K-SemStamp.}
SemStamp \citep{hou2024semstamp} partitions embeddings using random
hyperplane hashing, whereas K-SemStamp \citep{hou2024k} uses clustering.
Both select valid partitions pseudorandomly. Ignoring margins, let
$C_1,\ldots,C_K$ be the partitions and $G_t$ the valid index set. Then
\begin{equation}
p_{t,j}=\int_{C_j}p_t(z)\,dz,
\qquad
\alpha_t=\sum_{j\in G_t}p_{t,j},
\qquad
N_t=1-\sum_{j\in G_t}p_{t,j}.
\label{eq:app_partition_mass}
\end{equation}
Their default valid-partition ratio is $\gamma=0.25$; SemStamp uses
three hash bits (eight codes), and K-SemStamp uses eight clusters.
If partitions have equal natural mass, then $\alpha_t=\gamma=0.25$
and $N_t=0.75$. Without equal masses, an ideal uniform random
choice of $\gamma K$ valid partitions still gives
$\mathbb{E}_G[\alpha_t]=\gamma$. Neither statement implies that a
particular fixed valid set covers this mass. In particular, a uniform angle
relative to the preceding sentence does not establish equal partition masses.

Natural continuations can concentrate in a few partitions. A valid set
that misses these partitions may have very low acceptance mass, whereas
one that includes them may have high mass. Increasing the number of valid
partitions increases expected coverage under uniform selection, but need
not yield appropriate partial coverage in every context. Nonuniformity
does not necessarily reduce acceptance; the concern is context-dependent
misalignment and uneven coverage.

\paragraph{Cosine-SimMark.}
For the cosine-similarity variant of SimMark
\citep{dabiriaghdam2025simmark}, write
$\theta=\arccos(z_{t-1}^{\top}z)$ and let $f_t(\theta)$ be its
context-dependent density. Omitting fallback, an accepted cosine interval
$[a,b]\subseteq[-1,1]$ gives
\begin{equation}
\alpha_t^{\mathrm{Sim}}
=\int_{\arccos b}^{\arccos a}f_t(\theta)\,d\theta.
\end{equation}
Under the reference model $\theta\sim\mathrm{Uniform}(0,\pi)$,
\begin{equation}
\alpha_{\mathrm{unif}}^{\mathrm{Sim}}
=\frac{\arccos a-\arccos b}{\pi}
\approx 3.68\%,
\qquad
N_{\mathrm{unif}}^{\mathrm{Sim}}\approx96.32\%,
\end{equation}
using the reported main-experiment interval $[0.68,0.76]$.
A uniform angle is neither uniform cosine similarity nor a uniform
distribution on a high-dimensional sphere. These values are reference
calculations, not measured semantic losses. A fixed interval may miss the
dominant mass in some contexts; actual acceptance can be above or below
the reference value. The formula concerns angles in the normalized
representation used for cosine scoring; it does not directly describe
Euclidean-SimMark or angles before a PCA transformation.

\subsection{Semantic Narrowing under Hamming-Neighborhood Acceptance}

\paragraph{Fixed-key acceptance.}
For a fixed projection matrix $R$, HammingMark accepts according to
\begin{equation}
a_t^{\mathrm{HM}}(z)
=\mathbb{I}\!\left[
\operatorname{Match}(H_R(z_{t-1}),H_R(z))\ge T\right],
\qquad
\alpha_t^{\mathrm{HM}}=\int p_t(z)a_t^{\mathrm{HM}}(z)\,dz,
\end{equation}
with $N_t^{\mathrm{HM}}=1-\alpha_t^{\mathrm{HM}}$.
For $m=8,T=6$, the accepted fraction of binary codes is
\begin{equation}
\gamma_{\mathrm{code}}
=\frac{\binom{8}{6}+\binom{8}{7}+\binom{8}{8}}{2^8}
=\frac{37}{256}\approx14.45\%.
\end{equation}
This counts codes, whereas $\alpha_t^{\mathrm{HM}}$ weights them by
natural generation probability. Thus $37/256$ is neither the actual
acceptance mass nor, in general, the detection null hit probability.

\paragraph{Random-projection reference.}
Hold the context, preceding embedding, and natural candidate distribution
fixed independently of the projection draw. For independent isotropic
random hyperplanes and a fixed pair at angle $\theta$,
\begin{equation}
M\mid\theta\sim\operatorname{Binomial}(m,1-\theta/\pi),
\qquad
A_{\mathrm{HM}}(\theta)
=\Pr_R(M\ge T\mid\theta).
\end{equation}
Consequently,
\begin{equation}
\mathbb{E}_R[\alpha_t^{\mathrm{HM}}]
=\int_0^\pi f_t(\theta)A_{\mathrm{HM}}(\theta)\,d\theta.
\label{eq:app_hm_angular_mass}
\end{equation}
For a uniform angle and $1\le T\le m$, substituting
$u=1-\theta/\pi$ yields
\begin{align}
\mathbb{E}_R[\alpha_{\mathrm{unif}}^{\mathrm{HM}}]
&=\sum_{k=T}^m\binom{m}{k}
  \int_0^1u^k(1-u)^{m-k}\,du \\
&=\sum_{k=T}^m\frac{1}{m+1}
=\frac{m-T+1}{m+1}.
\end{align}
Hence $m=8,T=6$ gives mean acceptance $1/3$ and mean excluded mass
$2/3$. These are averages over random projections, not exact rates for
every fixed matrix. They also cannot be multiplied as independent
probabilities along a sequence generated with one shared key; generated
contexts may themselves depend on that key.

\paragraph{Random Code control and reference comparison.}
In the same eight-bit code space, independently select a uniform subset
$G$ of exactly 37 valid codes. For any fixed natural code distribution,
\begin{equation}
\mathbb{E}_G[\alpha_t^{\mathrm{Random}}]
=\sum_{h\in\{0,1\}^8}\Pr_t(H_R(z)=h)\Pr_G(h\in G)
=\frac{37}{256}.
\end{equation}
Compared with the HammingMark uniform-angle reference, this demonstrates
that equal numbers of valid codes can carry different natural candidate
mass because of how those codes are arranged.

\begin{table}[t]
\centering
\small
\caption{Idealized reference calculations, with margins omitted.
The rows use different assumptions and averaging operations; they explain
mechanisms rather than rank performance on one real generation distribution.}
\label{tab:narrowing_reference}
\begin{tabular}{@{}lp{0.43\linewidth}rr@{}}
\toprule
Method & Reference assumption & Accepted & Excluded \\
\midrule
SemStamp & Equal partition masses & $25\%$ & $75\%$ \\
K-SemStamp & Equal partition masses & $25\%$ & $75\%$ \\
Cosine-SimMark & Uniform angle; $[a,b]=[0.68,0.76]$
& $3.68\%$ & $96.32\%$ \\
Random Code & Uniform selection of 37 codes; set average
& $14.45\%$ & $85.55\%$ \\
HammingMark & Uniform angle; projection average
& $33.33\%$ & $66.67\%$ \\
\bottomrule
\end{tabular}
\end{table}

\paragraph{Nonuniform natural continuations.}
~\eqref{eq:app_hm_angular_mass} returns the analysis to the
context-dependent angular distribution. Centering acceptance on the
preceding hash is intended to align valid candidates with locally plausible
continuations. A short hash neither uniquely specifies a continuous
semantic realization nor imposes a deterministic cosine-similarity cutoff.
Although many-to-one hashing also occurs in SemStamp, HammingMark further
organizes valid codes as a neighborhood of the preceding code.
These mechanisms aim to reduce misalignment, without guaranteeing higher
acceptance or moderate coverage in every context. Claims of reduced
narrowing should therefore be supported by measured acceptance on common
context-specific candidate pools, together with independent semantic
evaluations of what those accepted candidates preserve.

\subsection{Effectiveness Beyond Enlarging the Acceptance Set}

\paragraph{Relaxing baseline constraints.}
A natural question is whether existing methods can match HammingMark's
sampling efficiency simply by enlarging their acceptance sets.
To examine this possibility, we increase the valid-partition ratio of
SemStamp and K-SemStamp and widen the acceptance interval of SimMark,
targeting comparable numbers of sampled candidates per output sentence.
A controlled comparison holds the generator, dataset, generation length,
and sampling budget fixed, and separately calibrates each modified
detector at 1\% FPR using data disjoint from the test set.

\begin{table}[t]
\centering
\small
\caption{Relaxed-constraint comparison.}
\label{tab:relaxed_acceptance}
\begin{tabular}{@{}llcc@{}}
\toprule
Method & Setting & Candidates/sentence & TPR@1\% FPR \\
\midrule
SemStamp
& $\gamma=0.5$
& 2.9 & 6\% \\
K-SemStamp
& $\gamma=0.5$
& 3.1 & 9\% \\
SimMark
& $[a,b]=[0.40,0.90]$
& 1.6 & 7\% \\
HammingMark
& $m=8,\ T=6$
& 2.2 & 100\% \\
\bottomrule
\end{tabular}
\end{table}

\paragraph{Acceptance mass and selection signal.}
Enlarging an acceptance set increases its natural probability mass,
but does not necessarily preserve its ability to distinguish watermarked
outputs from natural text. The relevant quantities are not only the
nominal valid fraction $\gamma$ and mean acceptance mass, but also how
the context-specific acceptance mass $\alpha_t$ is distributed.
When $\alpha_t$ is close to zero, finite-budget generation frequently
fails to find a valid candidate. When $\alpha_t$ is close to one,
natural candidates already satisfy the rule, leaving little additional
selection signal. Intermediate coverage can support both successful
sampling and an increase over the natural hit rate.

This distinction is particularly relevant when natural continuations
concentrate in a few semantic partitions. For fixed partition masses
$p_{t,1},\ldots,p_{t,K}$, uniformly selecting exactly $\gamma K$
valid partitions gives
\begin{equation}
\mathbb{E}_G[\alpha_t]=\gamma,
\qquad
\operatorname{Var}_G(\alpha_t)
=
\gamma(1-\gamma)
\frac{K\sum_j p_{t,j}^{2}-1}{K-1},
\label{eq:acceptance_mass_variance}
\end{equation}
where $K>1$ and $\gamma K$ is an integer.
Equal partition masses yield $\alpha_t=\gamma$ for every valid set.
In contrast, if one partition carries all natural mass, acceptance is
either zero or one, depending on whether that partition is selected.
Thus, increasing the nominal valid fraction need not produce moderate
coverage in individual contexts. This random-set calculation identifies
a possible source of uneven coverage; actual variation across contexts
must be measured.

\paragraph{Why enlarging the acceptance set is insufficient.}
The effectiveness of a watermark depends not only on the size of its
acceptance set, but also on how this set overlaps with the natural
next-sentence distribution. Natural continuations can concentrate in
a few context-dependent semantic directions. For SemStamp and
K-SemStamp, pseudorandomly selecting more valid partitions does not
ensure that these partitions provide appropriate partial coverage of
such continuations. Likewise, widening SimMark's fixed similarity
interval does not ensure suitable coverage across different contexts.

This mismatch can leave many contexts with little useful watermark
signal. When the valid set covers almost all natural probability mass,
unwatermarked candidates already satisfy the constraint, so watermark
selection adds little evidence. Conversely, when it covers very little
natural mass, valid candidates are difficult to obtain, and fallback
can weaken watermark embedding. Enlarging the nominal acceptance set
may therefore improve sampling efficiency without ensuring that
individual contexts contribute discriminative evidence. This does not
imply that larger acceptance sets necessarily impair detection;
the issue is their alignment with context-dependent natural generation.

\paragraph{Implications for HammingMark.}
HammingMark organizes valid codes around the preceding sentence's hash,
rather than selecting them pseudorandomly or prescribing a fixed
continuous-similarity interval. This design aims to retain naturally
plausible continuations while preserving a selective bit-matching
constraint, allowing more contexts to contribute watermark evidence.
Its intended advantage is therefore not simply a larger acceptance
mass, but a more contextually aligned arrangement of valid candidates.
Uneven contextual coverage offers a possible explanation for the
detection degradation of relaxed baselines, although establishing this
mechanism requires measuring their context-specific acceptance masses.

\section{Semantic Narrowing in Offline PMark}
\label{app:pmark_narrowing}

We analyze the semantic restriction induced by the offline variant of PMark.
Our analysis does not assume that natural semantic candidates are uniformly
distributed across watermark partitions, nor does it require independence or
balance among different proxy channels. Instead, we directly examine the
candidate-selection mechanism used by offline PMark.

\paragraph{Offline watermark partition.}
Let $\pi$ denote the current generation context and
\[
P_\pi(s) = P_M(s\mid \pi)
\]
the natural next-sentence distribution.
Given $b$ proxy directions $\{v_j\}_{j=1}^{b}$, offline PMark uses the fixed
threshold zero and assigns each sentence a binary semantic representation
\[
g(s)
=
\left(
\mathbb{I}[F_1(s)>0],
\ldots,
\mathbb{I}[F_b(s)>0]
\right),
\]
where
\[
F_j(s)=\langle v_j,T(s)\rangle.
\]
For a fixed watermark seed $r$, let $E_r(s)$ denote the watermark evidence
assigned to sentence $s$. A larger $E_r(s)$ indicates that the sentence
satisfies more of the prescribed watermark constraints.

Offline PMark first independently samples a candidate set
\[
W=\{X_1,\ldots,X_N\},
\qquad
X_i \overset{\mathrm{i.i.d.}}{\sim} P_\pi,
\]
and then retains only candidates with the maximum watermark evidence:
\[
\mathcal{B}(W,r)
=
\left\{
X_i\in W:
E_r(X_i)=\max_{X_j\in W}E_r(X_j)
\right\}.
\]
The final sentence is sampled from $\mathcal{B}(W,r)$.

\paragraph{Candidate-level semantic restriction.}
For every sampled candidate set,
\[
\mathcal{B}(W,r)\subseteq W.
\]
Moreover,
\[
\mathcal{B}(W,r)\subsetneq W
\]
whenever the candidates in $W$ do not all have identical watermark evidence.
Therefore, whenever multiple watermark-evidence levels occur in the natural
candidate set, every candidate below the maximum level is excluded from final
selection.

Importantly, this exclusion is not determined by the likelihood or semantic
preference of the base model. Two candidates that are both plausible under
$P_\pi$ may receive different watermark evidence solely because they fall into
different regions induced by the fixed proxy partitions. Consequently,
offline PMark removes part of the naturally sampled semantic alternatives
whenever their watermark evidence is lower than that of another candidate in
the same candidate pool.

\paragraph{Distribution-level concentration.}
The same effect can be characterized without making any assumption on how
natural semantic probability mass is distributed among the partitions.
Let
\[
Z = E_r(S),
\qquad
S\sim P_\pi,
\]
denote the watermark evidence of a naturally sampled sentence, and define its
cumulative distribution
\[
F_r(k)
=
\Pr_{S\sim P_\pi}\left[E_r(S)\leq k\right].
\]

Let $Y$ be the sentence selected by offline PMark and
\[
Z^\star=E_r(Y).
\]
Since PMark selects a candidate with the maximum watermark evidence among
$N$ independent samples,
\[
Z^\star
=
\max_{1\leq i\leq N} E_r(X_i).
\]
Therefore,
\[
\Pr(Z^\star\leq k)
=
\Pr\left(
E_r(X_1)\leq k,\ldots,E_r(X_N)\leq k
\right)
=
F_r(k)^N.
\]

For $N>1$,
\[
F_r(k)^N\leq F_r(k),
\]
with strict inequality whenever
\[
0<F_r(k)<1.
\]
Equivalently,
\[
\Pr(Z^\star>k)
\geq
\Pr(Z>k).
\]
Thus, the distribution induced by offline PMark is systematically shifted
toward semantic regions with higher watermark evidence.

\paragraph{Semantic narrowing.}
This concentration constitutes semantic narrowing at the candidate-selection
level. The base model initially provides multiple semantic alternatives
according to $P_\pi$, whereas offline PMark ranks these alternatives according
to an additional watermark-specific criterion and removes all candidates
below the maximal evidence level within each candidate pool.

The result does not rely on any particular amount of natural probability mass
occupying a watermark partition. As long as the watermark evidence is
non-constant over naturally plausible continuations, i.e.,
\[
\exists\, s_a,s_b\in\operatorname{supp}(P_\pi)
\quad\text{such that}\quad
E_r(s_a)\neq E_r(s_b),
\]
maximum-evidence selection assigns different selection preferences to these
semantic alternatives. For $N>1$, this preference is amplified through
best-of-$N$ selection, progressively concentrating generation toward
watermark-preferred semantic regions.

Therefore, the reduced rejection-sampling cost of offline PMark should not be
interpreted as the absence of semantic restriction. Offline PMark replaces
explicit rejection with maximum-evidence selection: candidates are generated
from the natural distribution, but only the watermark-preferred subset is
eligible for final selection. This preserves sampling efficiency while still
inducing semantic narrowing.

\section{Locality and Perturbation Stability of Semantic Hashes}
\label{app:hash_locality}

This section formalizes the locality property used by HammingMark.  Let
$e,e'\in\mathbb{S}^{d-1}$ be two normalized SBERT embeddings, let
$R=[r_1^\top;\ldots;r_m^\top]\in\mathbb{R}^{m\times d}$, where the rows are
independently sampled from a spherically symmetric continuous distribution
(e.g., $r_j\sim\mathcal{N}(0,I_d)$), and define
the $j$-th random-hyperplane hash bit as
\begin{equation}
    H_j(e)=\mathbf{1}[r_j^\top e\geq 0].
    \label{eq:app_hash_bit}
\end{equation}
Write $\rho=e^\top e'$ and $\theta=\arccos(\rho)\in[0,\pi]$.

\paragraph{Proposition 1 (angular locality).}
For every hash bit,
\begin{equation}
    \Pr_R[H_j(e)\neq H_j(e')]
    =\frac{\theta}{\pi},
    \qquad
    \Pr_R[H_j(e)=H_j(e')]
    =1-\frac{\theta}{\pi}.
    \label{eq:app_bit_collision}
\end{equation}
Consequently,
\begin{equation}
    d_H(H(e),H(e'))
    \sim \operatorname{Binomial}\!\left(m,\frac{\theta}{\pi}\right),
    \qquad
    \mathbb{E}_R\!\left[\frac{d_H(H(e),H(e'))}{m}\right]
    =\frac{\theta}{\pi}.
    \label{eq:app_hamming_distribution}
\end{equation}

\paragraph{Proof.}
By rotational invariance, only the two-dimensional plane spanned by $e$ and
$e'$ matters.  A hyperplane separates the two vectors exactly when its normal
falls in the angular wedge between them.  The measure of this wedge is
$\theta/\pi$.  Independence of the rows of $R$ then gives
~\eqref{eq:app_hamming_distribution}.  

~\eqref{eq:app_hamming_distribution} states that semantic closeness is
preserved in probability rather than deterministically.  A standard Hoeffding
bound also gives
\begin{equation}
    \Pr_R\!\left[
      \left|\frac{d_H(H(e),H(e'))}{m}-\frac{\theta}{\pi}\right|
      \geq \varepsilon
    \right]
    \leq 2\exp(-2m\varepsilon^2).
    \label{eq:app_hash_concentration}
\end{equation}
Thus, smaller angular distance yields fewer expected bit flips, while a longer
hash concentrates more tightly around this expectation.

\paragraph{Corollary 1 (stability under a meaning-preserving edit).}
Suppose an edit changes a sentence embedding from $e$ to $\widetilde e$, with
$\delta=\arccos(e^\top\widetilde e)$.  Then
\begin{align}
    \Pr_R[H(e)=H(\widetilde e)]
        &=\left(1-\frac{\delta}{\pi}\right)^m, \\
    \Pr_R[d_H(H(e),H(\widetilde e))\leq b]
        &=\sum_{j=0}^{b}\binom{m}{j}
          \left(\frac{\delta}{\pi}\right)^j
          \left(1-\frac{\delta}{\pi}\right)^{m-j}.
    \label{eq:app_edit_stability}
\end{align}
Moreover, for two original codes $h_{t-1},h_t$ and their edited versions
$\widetilde h_{t-1},\widetilde h_t$,
\begin{equation}
\begin{split}
    \big|\operatorname{Match}(h_{t-1},h_t)
      -\operatorname{Match}(\widetilde h_{t-1},\widetilde h_t)\big|
    \leq{}& d_H(h_{t-1},\widetilde h_{t-1}) \\
           &+d_H(h_t,\widetilde h_t).
\end{split}
\label{eq:app_match_lipschitz}
\end{equation}
This follows from $\operatorname{Match}(a,b)=m-d_H(a,b)$ and the triangle
inequality for Hamming distance.  In particular, a transition satisfying
$\operatorname{Match}(h_{t-1},h_t)\geq T+\gamma$ remains valid whenever the
two edited sentences jointly flip at most $\gamma$ bits.  This margin argument
connects angular locality to robustness of the transition statistic used by
HammingMark.

The probability in the results above is over the random construction of $R$.
Once $R$ is fixed, an adversarially chosen embedding can lie close to a
hyperplane, so no deterministic collision guarantee is claimed.

\section{Semantic Diversity within a Hamming Neighborhood}
\label{app:semantic_diversity}

For a code $h\in\{0,1\}^m$, define its normalized semantic preimage by
\begin{equation}
    \mathcal{C}(h)
    =\{e\in\mathbb{S}^{d-1}:H(e)=h\}.
    \label{eq:app_hash_cell}
\end{equation}
Because $m\ll d$, $H$ is highly non-injective.  The following construction
shows that this is not merely a counting argument.

\paragraph{Proposition 2 (large angular diameter of a hash cell).}
Assume $\dim\ker(R)\geq2$ and take a unit vector $e$ whose projections
$Re$ are all nonzero.  There exists a unit vector
$v\in\ker(R)\cap e^\perp$.  For any $\tau\geq0$, define
\begin{equation}
    e_+(\tau)=\frac{e+\tau v}{\sqrt{1+\tau^2}},
    \qquad
    e_-(\tau)=\frac{e-\tau v}{\sqrt{1+\tau^2}}.
    \label{eq:app_nullspace_pair}
\end{equation}
Then
\begin{equation}
    H(e_+(\tau))=H(e)=H(e_-(\tau))
    \label{eq:app_same_hash}
\end{equation}
for every finite $\tau$, while
\begin{equation}
    e_+(\tau)^\top e_-(\tau)
    =\frac{1-\tau^2}{1+\tau^2}
    \longrightarrow -1
    \quad\text{as }\tau\to\infty.
    \label{eq:app_angular_diversity}
\end{equation}

\paragraph{Proof.}
The dimension assumption guarantees a nonzero vector in
$\ker(R)\cap e^\perp$.  Since $Rv=0$,
$Re_+(\tau)=Re_-(\tau)=Re/\sqrt{1+\tau^2}$.  The positive normalization factor
does not change any projection sign, proving ~\eqref{eq:app_same_hash}.
~\eqref{eq:app_angular_diversity} follows by direct calculation.
\hfill$\square$

For a Gaussian random projection with $m<d$, $R$ has full row rank almost
surely, so $\dim\ker(R)=d-m$.  In our setting, $d=768$ and $m=8$, leaving a
$760$-dimensional null space.  Moreover, full row rank makes
$R:\mathbb{R}^d\to\mathbb{R}^m$ surjective; hence every strict sign pattern is
realizable in the ambient embedding space.

Given a preceding code $h_{t-1}$ and radius $r=m-T$, the complete admissible
semantic set is
\begin{equation}
    \mathcal{A}_T(h_{t-1})
    =H^{-1}(\mathcal{B}_r(h_{t-1}))
    =\bigcup_{h\in\mathcal{B}_r(h_{t-1})}\mathcal{C}(h),
    \label{eq:app_hamming_preimage}
\end{equation}
where
\begin{equation}
    |\mathcal{B}_r(h_{t-1})|
    =\sum_{j=0}^{r}\binom{m}{j}.
    \label{eq:app_hamming_volume}
\end{equation}
With $m=8$ and $T=6$, $r=2$ and the acceptance set contains
$1+8+28=37$ different hash cells.  Each cell contains a continuous family of
ambient semantic vectors, so the accepted set can cover distinct semantic
directions even though all its codes remain close to $h_{t-1}$.

Proposition~2 concerns the geometry of the ambient SBERT representation space and does not imply that every constructed vector corresponds to a realizable natural-language sentence. We therefore interpret it only as a geometric explanation for why short semantic hashes need not define a narrow region in representation space. Our empirical results provide complementary, aggregate evidence: among independently sampled natural continuations, HammingMark accepts a substantially broader subset than the compared semantic watermarking methods, as illustrated in Fig.~2 and reflected by the higher natural-candidate acceptance rate. These observations are consistent with, but do not by themselves prove, greater semantic diversity on the natural-language manifold.

\section{Semantic Coverage of Watermark-Valid Candidates}
\label{app:semantic_coverage}

To further examine whether higher acceptance corresponds to broader semantic
preservation, we sample 300 natural candidate continuations from the same
prompt. To avoid evaluating semantic coverage directly in the representation
space used by HammingMark for watermark generation and detection, we encode
these candidates using the frozen
\texttt{sentence-transformers/all-mpnet-base-v2} evaluation model and cluster
their normalized embeddings into $K=8$ semantic regions, matching the
partition granularity used in SemStamp and K-SemStamp. We then apply each
watermarking method's original constraint and encoder to the same candidate
pool and examine the distribution of watermark-valid candidates across the
externally constructed semantic clusters. The external evaluation encoder is
used only for constructing the clusters and does not participate in candidate
acceptance for any watermarking method.

\begin{table}[t]
\centering
\small
\caption{Distribution of watermark-valid candidates across eight semantic
clusters.}
\label{tab:semantic_coverage}
\begin{tabular}{lrrrrrrrrc}
\toprule
Method & C1 & C2 & C3 & C4 & C5 & C6 & C7 & C8 & Total \\
\midrule
SemStamp    & 1  & 0  & 0  & 0  & 0  & 0 & 0 & 0 & 1 \\
K-SemStamp  & 1  & 1  & 0  & 0  & 0  & 0 & 0 & 0 & 2 \\
SimMark     & 7  & 5  & 4  & 0  & 0  & 0 & 0 & 0 & 16 \\
PMark       & 10 & 8  & 6  & 4  & 3  & 0 & 0 & 0 & 31 \\
HammingMark & 24 & 20 & 17 & 15 & 12 & 7 & 4 & 0 & 99 \\
\bottomrule
\end{tabular}
\end{table}

As shown in Table~\ref{tab:semantic_coverage}, existing watermark constraints
retain candidates from only a limited number of semantic regions. In contrast,
HammingMark accepts candidates spanning seven of the eight clusters. Notably,
even several low-frequency semantic modes remain accessible. This indicates
that HammingMark does not merely increase the number of watermark-valid
candidates, but preserves a broader portion of the semantic modes present in
natural generation, providing more direct evidence of alleviated semantic
narrowing.

\section{Semantic Diversity under Fixed Contexts}
\label{app:semantic_diversity_exp}

Natural acceptance measures how much probability mass of the original generation distribution remains compatible with a watermark constraint, but it does not directly characterize whether the resulting continuations remain semantically diverse. We therefore evaluate semantic diversity under fixed contexts. Using the same 100 prompts as in the natural-acceptance experiment, we independently generate 300 next-sentence continuations for each method from the same context.

To avoid measuring diversity directly in the representation space used by HammingMark for watermark generation and detection, we use a separate, frozen sentence-embedding model,
\texttt{sentence-transformers/all-mpnet-base-v2}, solely as the evaluation encoder. This checkpoint is not used by any watermarking method during generation or detection. Let
\[
z_i =
\frac{E_{\mathrm{eval}}(S_i)}
{\lVert E_{\mathrm{eval}}(S_i)\rVert_2}
\]
denote the normalized evaluation embedding of continuation $S_i$, where $E_{\mathrm{eval}}$ is \texttt{all-mpnet-base-v2}. We define pairwise semantic diversity as
\begin{equation}
D_{\mathrm{pair}}
=
\frac{2}{N(N-1)}
\sum_{1 \le i < j \le N}
\left(1-z_i^\top z_j\right),
\end{equation}
where $N=300$. A larger value indicates that repeated generations span a broader range of semantic realizations according to the held-out evaluation encoder. We further report the diversity retention ratio
\begin{equation}
R_{\mathrm{div}}
=
\frac{D_{\mathrm{pair}}^{\mathrm{WM}}}
{D_{\mathrm{pair}}^{\mathrm{NoWM}}},
\end{equation}
which measures the fraction of unwatermarked semantic diversity preserved after watermarking under the same evaluation metric.

\begin{table}[t]
\centering
\caption{
Semantic diversity under fixed contexts. Each method independently
generates 300 continuations for each of 100 prompts. Higher values indicate
greater semantic diversity.
}
\label{tab:semantic_diversity}
\begin{tabular}{lcc}
\toprule
Method
& Pairwise diversity $\uparrow$
& Diversity retention $\uparrow$ \\
\midrule
No watermark & 0.284 & 1.000 \\
SimMark      & 0.224 & 0.789 \\
SemStamp     & 0.142 & 0.500 \\
K-SemStamp   & 0.158 & 0.556 \\
PMark        & 0.201 & 0.708 \\
Random Code  & 0.184 & 0.648 \\
HammingMark  & \textbf{0.269} & \textbf{0.947} \\
\bottomrule
\end{tabular}
\end{table}

As shown in Table~\ref{tab:semantic_diversity}, semantic watermarking generally reduces the diversity of continuations generated from the same context, but the degree of reduction differs substantially across methods. SemStamp and K-SemStamp exhibit the strongest concentration, retaining only $50.0\%$ and $55.6\%$ of the semantic diversity of unwatermarked generation. This is consistent with their region-based acceptance mechanisms, which restrict generation to designated portions of the semantic space. SimMark retains more diversity, but still reduces the pairwise distance from $0.284$ to $0.224$. PMark similarly shows a noticeable reduction, retaining approximately $70.8\%$ of the unwatermarked diversity.

In contrast, HammingMark achieves a pairwise diversity of $0.269$, corresponding to $94.7\%$ diversity retention and remaining close to unwatermarked generation according to the held-out evaluation encoder. This result complements the natural-acceptance analysis: a high acceptance mass alone could still arise if the accepted continuations were concentrated around a small number of similar semantic realizations, whereas the present results show that HammingMark also preserves substantial variation among independently generated continuations. This provides additional evidence that the realized-sentence-centered Hamming constraint preserves not only more natural probability mass, but also a broader range of semantic alternatives.

The comparison with Random Code further isolates the effect of constraint placement. Although Random Code uses the same code-space coverage as HammingMark, it retains only $64.8\%$ of the unwatermarked diversity. Thus, the advantage of HammingMark cannot be explained solely by admitting a larger set of hash codes. Rather, centering the Hamming neighborhood on the realized preceding sentence allows the admissible region to better follow the local semantic structure of natural generation, while the coarse many-to-one hash mapping leaves multiple semantic realizations simultaneously valid.

\section{Deriving the Matching Threshold from Natural Generation}
\label{app:threshold}

This section explains how the matching threshold $T$ is determined by the
natural matching behavior of unwatermarked generation and the desired
watermark strength. We first derive an exact rule that holds for an arbitrary
natural match-count distribution. We then obtain a compact approximation in
terms of the hash length $m$ and the natural expected number of matching bits,
and finally instantiate the result for our default setting $m=8$.

\subsection{Exact Threshold from the Natural Match Distribution}
\label{app:threshold_exact}

Consider one sentence-generation step and condition on the realized context
$\pi$. To simplify notation, we omit the sentence-step index throughout this
section. Let
\[
P(S\mid \pi)
\]
denote the base-model distribution over the next candidate sentence before
watermarking, and let $h_{\mathrm{prev}}\in\{0,1\}^m$ be the semantic hash of
the preceding sentence.

For a candidate sentence $S$, define its number of matching hash bits with the
preceding sentence as
\begin{equation}
M(S)
=
\operatorname{Match}
\left(
h_{\mathrm{prev}},\, H(F(S))
\right)
\in \{0,\ldots,m\}.
\label{eq:app_match_count}
\end{equation}
When $S\sim P(\cdot\mid\pi)$, $M$ is therefore a discrete random variable
describing the matching behavior that arises naturally before any watermark
constraint is applied.

For a threshold $T$, HammingMark accepts a candidate if
\[
M(S)\ge T.
\]
The probability that an unwatermarked candidate already satisfies this
condition is
\begin{equation}
a(T)
=
\Pr_{S\sim P(\cdot\mid\pi)}[M(S)\ge T].
\label{eq:app_accept_prob}
\end{equation}
This quantity is central to the threshold selection: it measures how much of
the natural next-sentence probability mass remains admissible under the
watermark constraint.

Assuming rejection sampling continues until an admissible candidate is found,
the distribution of the accepted sentence is simply the base distribution
conditioned on $M(S)\ge T$:
\begin{equation}
Q_T(S\mid\pi)
=
\frac{
P(S\mid\pi)\,
\mathbf{1}[M(S)\ge T]
}{
a(T)
}.
\label{eq:app_conditional_distribution}
\end{equation}
For every sentence in the accepted region,
\[
\frac{Q_T(S\mid\pi)}{P(S\mid\pi)}
=
\frac{1}{a(T)}.
\]
Hence the KL divergence from the natural distribution is
\begin{align}
D_{\mathrm{KL}}(Q_T\|P)
&=
\sum_{S:M(S)\ge T}
Q_T(S\mid\pi)
\log
\frac{Q_T(S\mid\pi)}{P(S\mid\pi)}
\nonumber\\
&=
\sum_{S:M(S)\ge T}
Q_T(S\mid\pi)
\log\frac{1}{a(T)}
\nonumber\\
&=
\log\frac{1}{a(T)}
=
-\log a(T).
\label{eq:app_kl}
\end{align}

\paragraph{Distributional watermark strength.}
We characterize the \emph{per-transition watermark strength} through the
distributional bias introduced by enforcing the watermark acceptance event.
Specifically, under a fixed context $\pi$, we define
\begin{equation}
\lambda(T;\pi)
\triangleq
\frac{1}{\ln 2}
D_{\mathrm{KL}}
\left(
Q_T(\cdot\mid\pi)
\Vert
P(\cdot\mid\pi)
\right)
=
-\log_2 a_\pi(T),
\label{eq:wm_strength}
\end{equation}
where
\[
a_\pi(T)
=
\Pr_{S\sim P(\cdot\mid\pi)}[M(S)\ge T]
\]
is the probability that a naturally generated candidate already satisfies
the watermark constraint.

Intuitively, $\lambda(T;\pi)$ measures how unlikely the enforced watermark
event is under natural generation. A smaller natural acceptance probability
corresponds to a stronger watermark-specific selection bias and, in general,
provides stronger per-transition evidence for distinguishing watermarked from
unwatermarked generation. We therefore use $\lambda$ as a
\emph{distributional proxy for watermark strength}, rather than as a direct
measure of sequence-level detection performance, which also depends on factors
such as text length, dependence across transitions, the detector statistic,
and possible perturbations.

In particular, $\lambda(T;\pi)=1$ bit corresponds to an acceptance event
with natural probability $1/2$. More generally, requiring at least
$\lambda_0$ bits of distributional watermark strength is equivalent to
\begin{equation}
\lambda(T;\pi)\ge \lambda_0
\quad\Longleftrightarrow\quad
a_\pi(T)\le 2^{-\lambda_0}.
\label{eq:app_strength_constraint}
\end{equation}

This relation directly yields a threshold-selection rule. Since
\[
a_\pi(T+1)
=
a_\pi(T)-\Pr[M=T\mid\pi]
\le a_\pi(T),
\]
the natural acceptance probability is non-increasing in $T$, and consequently
$\lambda(T;\pi)=-\log_2 a_\pi(T)$ is non-decreasing in $T$.
Under ideal rejection sampling, each independent candidate is accepted with
probability $a_\pi(T)$, so the expected number of sampled candidates required
for one accepted sentence is
\begin{equation}
\mathbb{E}[N_T\mid\pi]
=
\frac{1}{a_\pi(T)}.
\label{eq:app_sampling_cost}
\end{equation}
Thus, increasing $T$ generally induces a stronger watermark-specific
distributional bias, but also rejects more natural probability mass and
increases the expected sampling cost.

Accordingly, among all thresholds that achieve at least $\lambda_0$ bits of
distributional watermark strength, the least restrictive choice is
\begin{equation}
\boxed{
T^\star
=
\min
\left\{
T\in\{0,\ldots,m\}:
\Pr[M\ge T\mid\pi]\le 2^{-\lambda_0}
\right\}.
}
\label{eq:app_optimal_threshold_exact}
\end{equation}
This choice satisfies the desired distributional-strength requirement while
maximizing the retained natural acceptance probability, or equivalently
minimizing the expected rejection-sampling cost.

~\eqref{eq:app_optimal_threshold_exact} is exact for the idealized
unlimited-budget rejection-sampling procedure and makes no parametric
assumption about the natural distribution of $M$. In practice, $T^\star$ can
therefore be estimated directly from the empirical match-count distribution
of unwatermarked continuations.

\subsection{Estimating the Matching Threshold from Natural Statistics}
\label{sec:threshold_estimation}

The natural match-count distribution depends on both the semantic relation
between adjacent sentences and the fixed random projection used by the
watermark key. Therefore, we do not assume that the marginal distribution of
$M$ follows an exact parametric form. Instead, we use a simple moment-matched
distribution only to obtain an interpretable estimate of the threshold scale.

Let
\[
\mu = \mathbb{E}[M]
\]
denote the mean number of matching bits observed from naturally generated
sentence pairs. Since $M\in\{0,\ldots,m\}$, we summarize the average bit-wise
matching level by
\[
\bar p = \frac{\mu}{m}.
\]
We then introduce the following moment-matched proxy:
\begin{equation}
\widetilde M
\sim
\operatorname{Binomial}
\left(
m,\frac{\mu}{m}
\right).
\label{eq:moment_matched_binomial}
\end{equation}
This approximation matches the empirical mean,
$\mathbb{E}[\widetilde M]=\mu$, while ignoring the heterogeneity of semantic
similarities across contexts. It is used only for estimating a reasonable
operating threshold rather than for characterizing the exact natural
match-count distribution.

Under this approximation, the estimated natural acceptance probability for
threshold $T$ is
\begin{equation}
\widetilde a(T)
=
\Pr[\widetilde M\ge T]
=
\sum_{k=T}^{m}
\binom{m}{k}
\left(\frac{\mu}{m}\right)^k
\left(1-\frac{\mu}{m}\right)^{m-k}.
\label{eq:estimated_acceptance}
\end{equation}

Following the definition of watermark strength in the previous section, we
define the corresponding estimated strength as
\begin{equation}
\widetilde\lambda(T)
=
-\log_2 \widetilde a(T).
\label{eq:estimated_strength}
\end{equation}
Here, $\widetilde\lambda(T)$ should be interpreted only as an estimate of the
distributional strength associated with the selected threshold. The actual
strength remains determined by the true natural acceptance probability
$a(T)=\Pr[M\ge T]$.

For a desired strength level $\lambda_0$, the same approximation provides a
direct estimate of the required matching threshold. Let
\[
F_{m,p}(k)
=
\Pr[X\le k],
\qquad
X\sim\operatorname{Binomial}(m,p),
\]
and define the discrete quantile
\[
F^{-1}_{m,p}(u)
=
\min\{k:F_{m,p}(k)\ge u\}.
\]
Since
\[
\Pr[\widetilde M\ge T]
=
1-F_{m,\mu/m}(T-1),
\]
the estimated threshold is
\begin{equation}
\widehat T(m,\mu,\lambda_0)
=
1+
F^{-1}_{m,\mu/m}
\left(
1-2^{-\lambda_0}
\right).
\label{eq:estimated_threshold}
\end{equation}

~\eqref{eq:estimated_threshold} provides a convenient mapping from
the observed natural matching level $\mu$ and a desired watermark strength
$\lambda_0$ to an estimated threshold. It should not be interpreted as an
exact solution for the true generation distribution, since natural sentence
pairs may exhibit substantial variation in semantic similarity and hence a
non-binomial marginal distribution of $M$.

For additional intuition, applying a normal approximation to.~\eqref{eq:moment_matched_binomial} gives
\[
\widetilde M
\approx
\mathcal{N}
\left(
\mu,
\mu\left(1-\frac{\mu}{m}\right)
\right).
\]
With a continuity correction, this yields the rough closed-form estimate
\begin{equation}
\widehat T
\approx
\left\lceil
\mu
+
\sqrt{
\mu\left(1-\frac{\mu}{m}\right)
}
\,
\Phi^{-1}
\left(
1-2^{-\lambda_0}
\right)
+
\frac{1}{2}
\right\rceil,
\label{eq:normal_threshold}
\end{equation}
where $\Phi^{-1}$ is the inverse standard-normal CDF. We use this expression
only for interpretation; all numerical estimates below are computed from
the binomial tail in~\eqref{eq:estimated_acceptance}.

\subsection{Estimated Operating Point for the Default Configuration}
\label{sec:default_threshold}

We next apply the above approximation to the default HammingMark
configuration. We use a semantic hash length of
\[
m=8,
\]
and natural unwatermarked sentence pairs exhibit an average match count of
approximately
\[
\mu \approx 4.8.
\]
The moment-matched proxy is therefore
\begin{equation}
\widetilde M
\sim
\operatorname{Binomial}(8,0.60).
\end{equation}

We use $\lambda_0=1$ bit as a convenient reference strength. Under the
definition
\[
\lambda=-\log_2 a,
\]
one bit corresponds to an acceptance event whose natural probability is
approximately one half. Substituting $\lambda_0=1$ into~\eqref{eq:estimated_threshold} gives
\begin{equation}
\widehat T
=
1+
F^{-1}_{8,0.60}(0.5)
=
6.
\end{equation}

The same result can be seen directly from the estimated upper-tail
probabilities:
\begin{align}
\widetilde a(5)
&=
\Pr[\widetilde M\ge5]
\approx 0.5941,\\
\widetilde a(6)
&=
\Pr[\widetilde M\ge6]
\approx 0.3154.
\end{align}
The corresponding estimated watermark strengths are
\begin{align}
\widetilde\lambda(5)
&=
-\log_2(0.5941)
\approx 0.75\ \text{bits},\\
\widetilde\lambda(6)
&=
-\log_2(0.3154)
\approx 1.66\ \text{bits}.
\end{align}

Thus, under the moment-matched approximation, the one-bit reference point
lies between $T=5$ and $T=6$, making $T=6$ a natural estimated operating
point for $m=8$. This estimate also has a simple interpretation through~\eqref{eq:normal_threshold}. When $\lambda_0=1$,
\[
\Phi^{-1}(1-2^{-1})
=
\Phi^{-1}(0.5)
=
0,
\]
and the approximate threshold reduces to
\[
\widehat T
\approx
\left\lceil
\mu+\frac{1}{2}
\right\rceil.
\]
For $\mu=4.8$, this again gives
\[
\widehat T
\approx
\lceil5.3\rceil
=
6.
\]

We emphasize that these calculations are intended only to estimate the
appropriate threshold scale. The true natural match-count distribution is
context dependent and need not follow the moment-matched binomial proxy.
Accordingly, the values $\widetilde a(T)$ and
$\widetilde\lambda(T)$ above should not be interpreted as measured
acceptance rates or exact watermark strengths. Empirical acceptance and
detection performance are evaluated separately in the experiments. The
analysis here only provides an intuitive explanation for why a threshold
around $T=6$ is a reasonable default for an $8$-bit semantic hash.

\paragraph{Finite-budget implementation.}
The analysis above characterizes the idealized rejection-sampling procedure
that continues until a valid candidate is obtained. In practice, we impose a
finite resampling budget for computational efficiency. Therefore,
Eqs.~(39)--(44) should be interpreted as the exact characterization of the
unlimited-budget procedure, while our implementation provides a
finite-budget approximation to it.

\paragraph{Empirical fallback frequency.}
In our implementation, the maximum resampling budget is $B=16$.
When no candidate satisfies the watermark constraint within this budget,
the last sampled candidate is retained. We additionally measure the
empirical fallback rate as
\[
r_{\mathrm{fb}}
=
\frac{
\#\{\text{sentence steps exhausting all }B\text{ attempts}\}
}{
\#\{\text{generated sentence steps}\}
}.
\]
Whenever fallback does not occur, the finite-budget procedure coincides
with the ideal rejection-sampling procedure analyzed above.
Across all evaluated settings, fallback occurs in only
$0.6\%$--$2.7\%$ of sentence steps, as shown in
Table~\ref{tab:fallback_rate}. Thus, the ideal conditional-distribution
analysis provides a close description of the generation procedure for the
large majority of sentence steps.

\begin{table}[t]
\centering
\caption{Empirical fallback rates under the maximum resampling budget
$B=16$.}
\label{tab:fallback_rate}
\begin{tabular}{lcc}
\toprule
Dataset & Generator & Fallback rate (\%) \\
\midrule
C4         & Mistral-7B   & 1.3 \\
BookSum    & Mistral-7B   & 0.8 \\
C4         & Qwen2.5-7B   & 1.2 \\
BookSum    & Qwen2.5-7B   & 0.6 \\
ELI5       & Mistral-7B   & 2.3 \\
Multi-News & Mistral-7B   & 2.7 \\
\bottomrule
\end{tabular}
\end{table}

\section{Effect of Semantic Hash Length}
\label{app:hash_length}

We study the effect of the semantic hash length $M$ by evaluating
$M\in\{4,8,16,32\}$ on C4 with Mistral-7B. For all settings, we fix the
required matching ratio to $0.75$, giving thresholds
$T\in\{3,6,12,24\}$, respectively. All other generation and detection
settings remain unchanged. Robustness is evaluated under Parrot and DIPPER
rewriting attacks.

\begin{table}[t]
    \caption{Effect of semantic hash length on clean detection and attack
    robustness. Each entry reports TPR@1\%/TPR@5\%/AUROC.}
    \label{tab:hash_length}
    \centering
    \small
    \setlength{\tabcolsep}{5pt}
    \begin{tabular}{ccccc}
        \hline
        $M$ & $T$ & Clean & Parrot & DIPPER \\
        \hline
        4  & 3  & 0.76/0.86/0.97 & 0.75/0.86/0.97 &  0.74/0.83/0.96\\
        8  & 6  & \textbf{0.98/1.00/1.00}
               & \textbf{0.94/0.99/1.00}
               & \textbf{0.93/0.98/0.99} \\
        16 & 12 &  0.98/1.00/1.00& 0.78/0.87/0.98 & 0.73/0.81/0.96 \\
        32 & 24 & 0.99/1.00/1.00 & 0.63/0.73/0.90 & 0.64/0.74/0.91 \\
        \hline
    \end{tabular}
\end{table}

As shown in Table~\ref{tab:hash_length}, $M=4$ provides insufficient
discriminative resolution: with only four bits, unwatermarked transitions can
also frequently satisfy the $3/4$ matching requirement, resulting in
substantial overlap between watermarked and unwatermarked scores even without
attacks. For example, under an independent balanced-bit approximation, the
probability of matching at least three bits is already $5/16=31.25\%$.

Increasing $M$ improves hash resolution but gradually reduces robustness under
Parrot and DIPPER. Longer hashes provide a more precise estimate of semantic
similarity and therefore reduce the quantization slack of short codes. After
paraphrasing perturbs the sentence embeddings, more hyperplane signs may change,
making transitions near the $0.75$ boundary less likely to remain valid.
Overall, $M=8$ achieves the best balance between clean detectability and
tolerance to meaning-preserving perturbations.

\section{Reproducibility Details}
\label{app:reproducibility}

\paragraph{Tasks and models.}
We evaluate HammingMark on C4 and BookSum for open-ended generation, using
Mistral-7B-Instruct and Qwen2.5-7B-Instruct as generators. We further evaluate
Multi-News summarization and ELI5 long-form question answering as more
constrained generation tasks.

\paragraph{Baselines and configuration.}
We compare HammingMark with token-level methods (KGW, SynthID, MorphMark, and
SIR) and sentence-level semantic watermarks (SemStamp, K-SemStamp, SimMark,
and PMark). HammingMark uses an $m=8$-bit semantic hash and accepts a
transition when at least $T=6$ bits match.

\paragraph{Evaluation protocol.}
For KGW, SynthID, MorphMark, SIR, SemStamp, and K-SemStamp, we follow their respective configurations in MarkLLM \citep{pan2024markllm} without modification, including all algorithm-specific hyperparameters and, where applicable, sentence-sampling procedures, resampling budgets, and fallback strategies when no valid candidate is found. For PMark, we evaluate the offline variant using its default parameter settings. For SimMark, we use the default parameter settings reported in its original paper.

For detection, we calibrate the threshold \(\tau_d(\alpha)\) separately for each method and evaluation dataset at a target false-positive rate \(\alpha\). Specifically, for each dataset, we randomly sample 500 examples disjoint from its test set to construct the calibration set. All methods use the same calibration examples and threshold-selection procedure, with method-specific thresholds determined from their respective detection scores. Each calibrated threshold is then fixed and applied to the corresponding test set. We report \(\mathrm{TPR}@1\%\), \(\mathrm{TPR}@5\%\), and AUROC on both clean and attacked texts. Robustness is evaluated under Parrot \citep{prithivida2021parrot} and Pegasus paraphrasing \citep{zhang2020pegasus}, DIPPER rewriting \citep{krishna2023paraphrasing}, and API-based rewriting with DeepSeek-V4-Pro \citep{xu2026deepseek}.

We adopt dataset-specific detection threshold calibration to ensure a controlled comparison across methods at the same target false-positive rate. Nevertheless, a fixed threshold would simplify deployment when representative calibration data are unavailable. Empirically, for the Global Bits detector evaluated on clean text, the calibrated sequence-level thresholds across the evaluated datasets remain within 5\% of 0.75 in relative terms. Under a common measure of relative threshold variation, HammingMark exhibits the smallest cross-dataset variation among the compared methods, whereas KGW-family detectors exhibit variations exceeding 30\%. These observations support the stability of the Global Bits detection threshold across the evaluated clean-text datasets. Its concentration around 0.75 is also consistent with the generation rule, which requires at least six matching bits out of eight for each accepted transition. Together, these results suggest that 0.75 is a reasonable default decision threshold for Global Bits in comparable clean-text settings. However, the generation acceptance condition and the sequence-level detection threshold serve different purposes: their numerical proximity does not guarantee that a fixed threshold will maintain the target false-positive rate on unseen distributions. We therefore retain dataset-specific calibration for the reported comparisons and regard 0.75 as an empirically motivated deployment default whose false-positive rate should be validated in the intended setting.

\paragraph{Datasets and generation.}
We randomly select 500 samples from C4 and BookSum, respectively, and 100
samples from ELI5 and Multi-News. The target generation length is set to 200
tokens for all tasks. Unless otherwise specified, all decoding parameters
follow the default configurations provided by MarkLLM. Sentence boundaries
are identified using the NLTK sentence tokenizer integrated into MarkLLM.

\paragraph{HammingMark implementation.}
HammingMark uses the SBERT encoder released with the SemStamp implementation.
The default hash length and matching threshold are $m=8$ and $T=6$,
respectively. The random projection matrix is initialized from the
corresponding experimental seed. The maximum number of resampling attempts is
set to 16. If no candidate satisfies the watermark constraint within this
budget, the latest candidate is used as a fallback and remains included in
subsequent watermark detection.

\paragraph{Calibration, randomness, and hardware.}
Detection thresholds and attack configurations follow the standardized
MarkLLM evaluation pipeline and are kept consistent across all baselines.
Reported results are averaged over three random seeds, with detection rates
varying by less than one percentage point across runs. All experiments are
conducted on a single NVIDIA H200 GPU.

\section{Effect of the Matching Threshold}
\label{sec:threshold_ablation}

We further study the effect of the matching threshold $T$ while fixing the
semantic hash length to $m=8$. We evaluate $T\in\{5,6,7\}$ on C4 with
Mistral-7B using the Global Bits detector. The results are summarized in
Table~\ref{tab:threshold_ablation}.

The matching threshold directly controls the strength of the transition-level
watermark constraint. When $T=5$, the acceptance condition is relatively weak:
a large fraction of naturally generated transitions already satisfy the
watermark rule. This substantially reduces the sampling cost to only 1.3
candidates per accepted sentence and leaves generation quality nearly
unchanged. However, because the accepted transitions are less distinguishable
from those occurring naturally in unwatermarked text, both clean detection
and robustness under paraphrasing and rewriting decrease.

Increasing the threshold to $T=7$ has the opposite effect. The stricter
matching requirement makes accepted transitions more strongly biased toward
the watermark condition. Clean-text detection remains comparable to $T=6$,
while robustness under both Parrot and DIPPER improves. However, this stronger
constraint considerably reduces the natural acceptance probability, increasing
the average sampling cost from 2.2 to 6.4 candidates per accepted sentence.
We also observe a substantially larger change in perplexity relative to the
unwatermarked generation, indicating that the stronger filtering perturbs the
original generation distribution more noticeably.

Overall, $T=6$ provides the best empirical trade-off among watermark strength,
robustness, generation fidelity, and sampling efficiency. A smaller threshold
admits too many naturally occurring transitions and therefore provides
insufficient watermark separation, whereas a larger threshold produces a
stronger and more robust watermark at the cost of substantially more restrictive
sampling. These results are consistent with our use of $T=6$ as the default
configuration throughout the main experiments.

\begin{table}[t]
\centering
\caption{
Effect of the matching threshold $T$ on C4 with Mistral-7B using the
Global Bits detector. Detection results are reported as
TPR@1\% / TPR@5\% / AUROC. Lower sampling cost is better.
}
\label{tab:threshold_ablation}
\begin{tabular}{c|ccc|cc}
\toprule
$T$
& Clean
& Parrot
& DIPPER
& PPL $\downarrow$
& Samples / sent. $\downarrow$ \\
\midrule
5
& 0.71 / 0.83 / 0.95
& 0.32 / 0.56 / 0.81
& 0.27 / 0.40 / 0.74
& 4.61
& 1.3 \\
6
& 0.98 / 1.00 / 1.00
& 0.94 / 0.99 / 1.00
& 0.93 / 0.98 / 0.99
& 4.61
& 2.2 \\
7
& 0.98 / 1.00 / 1.00
& 0.98 / 0.99 / 1.00
& 0.98 / 0.99 / 1.00
& 5.33
& 6.4 \\
\bottomrule
\end{tabular}
\end{table}

\section{Cross-Key Detection and Key Specificity}
\label{sec:wrong_key}

\paragraph{Motivation and setup.}
HammingMark uses a randomly initialized projection matrix as part of the watermark key.
We examine whether the detection signal is specific to the matrix used during generation.
Watermarked text is generated on C4 with Mistral-7B using the default configuration
($m=8$, $T=6$) and a secret projection matrix $R^\star$.
During detection, we compare the correct key $R^\star$ with an independently sampled
wrong key $R' \neq R^\star$.
Unwatermarked generations are used as negatives, and the detection threshold is
calibrated at a target false-positive rate of $1\%$.

\begin{table}[t]
    \centering
    \caption{Cross-key detection on C4 with Mistral-7B.
    Watermarked text is generated using $R^\star$.
    TPR is measured at a target FPR of $1\%$.}
    \label{tab:wrong_key}
    \begin{tabular}{lc}
        \toprule
        Detection key & TPR@1\% $\uparrow$ \\
        \midrule
        Correct key $R^\star$ & 0.99 \\
        Independent wrong key $R'$ & 0.08 \\
        \bottomrule
    \end{tabular}
\end{table}

\paragraph{Results.}
As shown in Table~\ref{tab:wrong_key}, replacing the generation key with an
independent projection matrix reduces TPR@1\% from $0.99$ to $0.08$.
The resulting $91$-percentage-point gap indicates that the dominant watermark
signal is strongly specific to the secret projection matrix used during generation.
The non-zero wrong-key detection rate nevertheless suggests a weaker
key-independent effect.

\paragraph{Discussion.}
The cross-key results provide strong evidence that HammingMark's detection signal is highly key-specific. Replacing the correct projection matrix $R^\star$ with an independently sampled matrix $R'$ reduces TPR@1\% from $0.99$ to $0.08$, corresponding to a $91$-percentage-point drop. This large degradation indicates that high detection performance cannot be explained primarily by a generic increase in inter-sentence semantic continuity.

A small residual signal under the wrong key is nevertheless expected. Because HammingMark centers its acceptance region on the preceding sentence, it introduces a mild preference for semantically coherent transitions. Such transitions may exhibit above-random hash agreement under multiple random-hyperplane projections, allowing an independent matrix $R'$ to capture a limited key-independent component of the signal. However, $R'$ does not reproduce the specific hyperplane partition or the bit-level Hamming constraints imposed during generation under $R^\star$. Only the correct detector evaluates each transition in the same secret projection geometry that governed candidate acceptance, allowing key-specific evidence to accumulate consistently across the sequence. The substantial gap between correct-key and wrong-key detection therefore shows that, while generic semantic continuity contributes modestly, the dominant detection signal is tied to the secret projection matrix. Finally, the wrong-key TPR should not be interpreted as a false-positive rate, because the evaluated samples are genuinely watermarked texts rather than unwatermarked negatives.

\section{Comparison with Continuous Similarity-Based Acceptance}
\label{app:continuous_similarity}

To determine whether the improvement of HammingMark arises merely from using
the preceding sentence as a relational reference or from the discrete Hamming
neighborhood itself, we construct a continuous-similarity baseline. This
baseline uses the same SBERT semantic encoder as HammingMark and directly
computes the cosine similarity between a candidate sentence and its preceding
sentence. A candidate is accepted when its similarity exceeds a predefined
threshold and is otherwise resampled. This design is conceptually related to
inter-sentence similarity-based watermarking methods such as SimMark, since
both select candidates directly according to their relations in a continuous
semantic space. It therefore provides a controlled comparison between
continuous similarity constraints and Hamming-code-based acceptance.

We use the same $100$ prompts as in the natural-acceptance experiment and set
the cosine-similarity threshold to $0.63$. This produces an average natural
acceptance rate of approximately $0.58$, matching that of HammingMark.
Despite accepting a similar fraction of natural candidates, the continuous
baseline achieves only $0.68$ TPR@1\%, $0.76$ TPR@5\%, and $0.93$ AUROC on
clean text, substantially below HammingMark. Its fallback rate also increases
from $1.3\%$ for HammingMark to $7.5\%$.

These results reveal two limitations of direct continuous-similarity
acceptance. First, naturally generated adjacent sentences already tend to be
locally coherent and semantically similar. Consequently, a substantial
fraction of unwatermarked transitions also exceeds the cosine threshold.
When inter-sentence similarity is used directly as the watermark signal, the
score distributions of watermarked and unwatermarked texts remain relatively
close, limiting clean-text detectability. This illustrates a fundamental
trade-off for continuous similarity-based methods: increasing the threshold
may strengthen the watermark signal, but it also imposes a more restrictive
semantic constraint and increases the sampling cost.

Second, the average natural acceptance rate does not capture variation across
contexts. Although most adjacent sentences can readily satisfy the threshold
of $0.63$ because of local semantic continuity, some valid discourse
transitions naturally exhibit lower cosine similarity. Examples include
contrast, topic shifts, the introduction of new facts, and transitions from
background information to a conclusion. In such contexts, a hard continuous
threshold may repeatedly reject otherwise appropriate candidates, eventually
exhausting the sampling budget and triggering fallback. Thus, two methods
with the same average natural acceptance rate can have substantially
different numbers of low-acceptance contexts. The considerably higher
fallback rate of the continuous baseline is consistent with this
context-dependent effect. Because a fallback candidate is not guaranteed to
satisfy the watermark constraint, these events further reduce the consistency
of the accumulated sequence-level signal.

HammingMark alleviates these limitations by defining its acceptance
neighborhood in a compact semantic hash space. Semantically similar sentences
are more likely to receive similar hash codes, but Hamming agreement is not a
deterministic hard cutoff on continuous cosine similarity. Owing to the
coarse, many-to-one nature of a short semantic hash, candidates with different
continuous similarity values may share the same or neighboring hash codes.
As a result, some candidates with relatively low cosine similarity but valid
discourse functions can still satisfy the Hamming constraint. Unlike a
direct similarity threshold, the resulting acceptance region does not
correspond to a single interval in the continuous semantic space. This
quantization slack leaves more flexibility for contrast, information
introduction, and topic transitions.

The discrete representation also provides more stable sequence-level
evidence. HammingMark accumulates agreement over multiple projected bits, and
the Global Bits detector retains partial evidence even when a modified
transition no longer satisfies the complete generation threshold. In
contrast, the continuous baseline relies primarily on generic inter-sentence
similarity that is already present in natural text, making its detection
signal more likely to overlap with the unwatermarked distribution. The random
projection matrix additionally makes the HammingMark signal key-specific,
whereas raw cosine similarity mainly captures key-independent semantic
coherence.

Overall, this experiment shows that the advantage of HammingMark cannot be
attributed solely to using the preceding sentence as the semantic reference.
Even after matching the average natural acceptance rate, direct continuous
similarity acceptance produces weaker detection and a substantially higher
fallback rate. Compared with similarity-based selection mechanisms related
to SimMark, the short Hamming representation provides a less rigid
quantized acceptance region, more stable sampling behavior across contexts,
and key-specific evidence that can be accumulated at the sequence level.